\documentclass{jfm}
\usepackage{latexsym}
\usepackage{graphics}
\usepackage{graphicx}
\usepackage{amsmath}
\usepackage{amssymb}
\usepackage{color}
\usepackage{natbib}
\usepackage{cellspace}
\usepackage{version}
\usepackage{array,multirow}

\def\Qadim{{Q}^*}

\def\rhop{\rho_{\mbox{\small p}}}
\def\rhop{\rho_p}

\newcommand{\beqa}{\begin{eqnarray}}
\newcommand{\beq}{\begin{equation}}
\newcommand{\ba}{\begin{array}}
\newcommand{\eeqa}{\end{eqnarray}}
\newcommand{\eeq}{\end{equation}}
\newcommand{\ea}{\end{array}}

\newcommand{\revPR}{\textcolor{black}}
\newcommand{\revPRb}{\textcolor{black}}

\title{Granular surface flows confined between flat, frictional walls. Part I: kinematics}

\author{Patrick Richard$^{1}$,
Alexandre Valance$^{2}$, 
Renaud Delannay$^{2}$ and 
Philippe Boltenhagen$^{2}$}
\affiliation{
$^{1}$ Univ Gustave Eiffel, MAST-GPEM, F-44344 Bouguenais, France\\
$^2$ IPR, Universit\'e de Rennes 1, CNRS UMR 6251, Campus Beaulieu, F-35042 Rennes, France}

\shorttitle{Granular flows confined between flat, frictional walls}
\shortauthor{P. Richard et al.}
\date{?? and in revised form ??}

\begin{document}

\maketitle

\begin{abstract}
We report and analyse the results of 
extensive discrete element method simulations  of three-dimensional gravity driven flows of cohesionless granular media  over an erodible bed, the whole being  confined between two flat and frictional sidewalls.
We focus on the role of sidewalls by performing simulations for different gap widths ($W$) between the two confining sidewalls: from $5$ to $30$ grain sizes ($d$). 
Our results indicate the existence of two distinct regimes: regime I for small flow angles $\theta$ ($\theta < \theta_c$ with $\theta_c\approx40^\circ$) and regime II at important flow angles ($\theta > \theta_c$). Regime I corresponds to dense flows whereas flows belonging to regime II exhibit a strong variation of the volume fraction through the depth. 
Three relevant lengths are identified in the system: $W$ the gap between sidewalls, $l$ the length characterizing the vertical variation of the volume fraction and $h$ a characteristic length associated with the vertical variation of the streamwise velocity.
Using these  lengths we can rescale the profiles of various flow properties ({e.g.} streamwise velocity, granular temperature, particle rotation\ldots). In regime II, in contrast to regime I, $l$ and $h$ have a similar behaviour. As a consequence, the rescaled profiles in regime II only involve  $h$ (or equivalently $l$) and $W$.  
Other dissimilarities exist between regimes I and II. In particular, the scaling of the flow rate with $h$ (at fixed $W$) differs in the two regimes, although they display a similar scaling with $W$ (at fixed flow angle).    
\end{abstract}

\section{Introduction: State of the Art}

\begin{figure}
\begin{center}
\includegraphics[width=0.85\columnwidth]{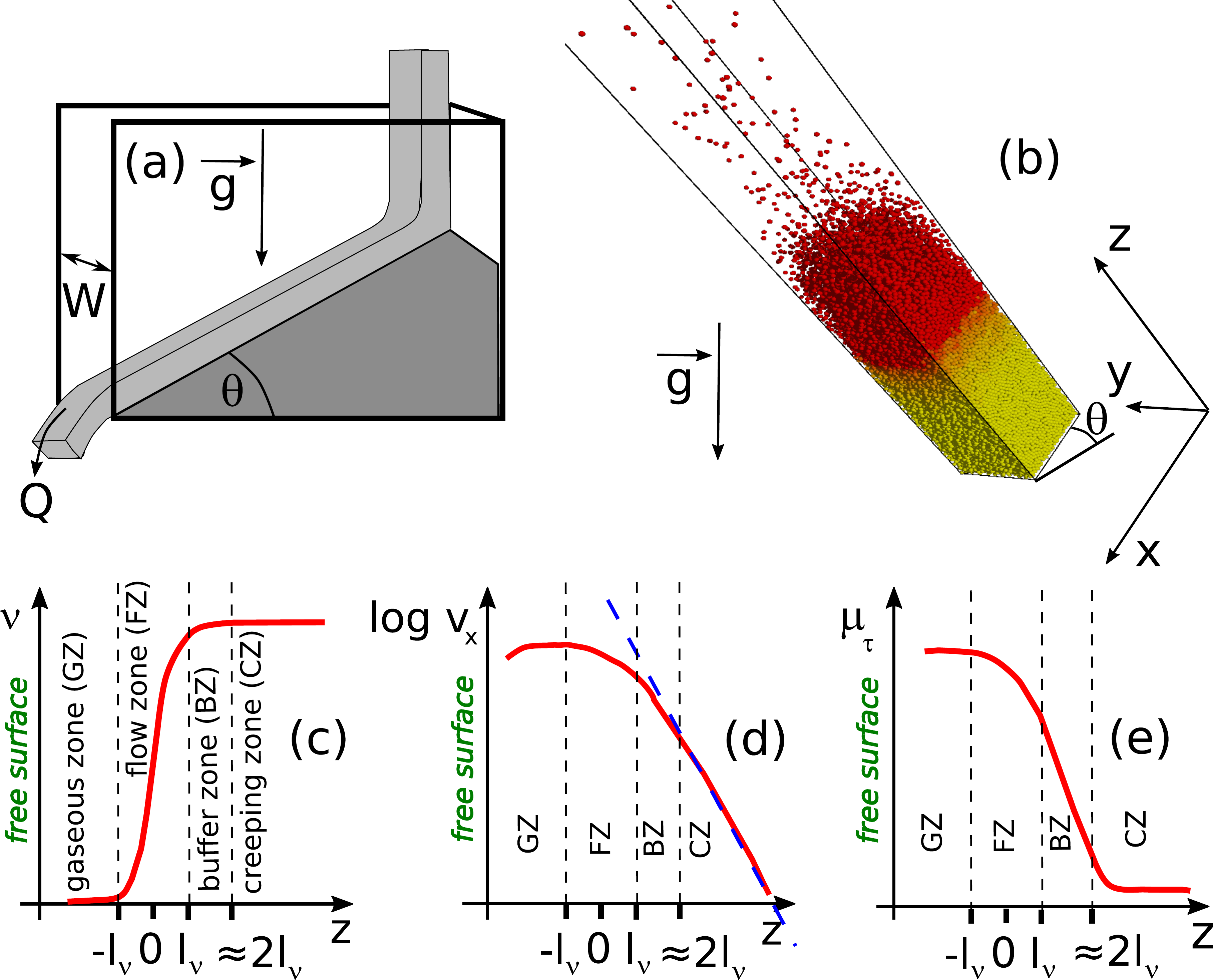}
\caption{
Sketch of the granular chute flow set-up: a surface flow occurs over a pile and is confined between two sidewalls separated by a gap $W$ (a). The angle of the flow, $\theta$, increases with the flow rate $Q$.
Typical three-dimensional snapshot for chute flow: $W/d=20$, $N=24,000$ (b). 
Flow is directed
down the incline along the $x$-axis and its angle is $\theta=45^\circ$. The two sidewalls are parallel to the $(xz)$ plane.
The axes $x$, $y$ and $z$ are also reported and the origin is arbitrary. Sketches of the vertical profiles of the volume fraction $\nu$ (c), logarithm of the streamwise velocity $v_x$ (d) and sidewall friction  $\mu_\tau$ (e) obtained in~\cite{Richard2008}. 
The top and the bottom of the system correspond to
$z\rightarrow -\infty$ and $z\rightarrow +\infty$, respectively, and $z=0$ is defined as the depth for which the volume fraction is equal to the half of its maximal value. The variation of the volume fraction is characterized by a length $l_\nu$ which allows us to define four zones from top  to bottom: (i) a dilute gaseous zone (GZ) located between $-\infty$ and $-l_\nu$ where particles have ballistic trajectories and experience rare collisions, (ii) a flowing zone (FZ) between $-l_\nu$ and $l_\nu$ where $\nu$ rapidly increases with depth, (iii) a buffer zone (BZ) between $l_\nu$ and approximately $2 l_\nu$  and finally (iv), below the buffer layer, a creeping zone (CZ) where $\nu = \nu_0 \simeq 0.6$.
}\label{fig:resume}
\end{center}
\end{figure}

Gravity driven granular flows are important, as they constitute a paradigm for the modelling of usual flows in industrial applications and geophysical processes. Significant progress has been made during the last decades in describing these flows. Nevertheless, a full understanding of their behaviour is far from being attained and their study remains an active field of research.
The nature of the boundary conditions at the bottom exerts a strong influence on dry granular flows driven by gravity~\citep{Delannay2007}. Most of the published studies are devoted to flows over a bumpy rigid base. Flows over an erodible base (\revPR{see Figure~\ref{fig:resume}a}) are nevertheless important to understand the dynamics of many dense flows in geophysical contexts such as landslides that move on deposits made of the same material that composes the flow (see \citep{Lemieux2000, Komatsu_PRL_2001, Khakhar2001, Taberlet2003, Jop_JFM_2005, Richard2008, Li_GranularMatter_2020} and references herein).
Lateral boundary can also exert a significant influence on the flow. Recently, several works have been devoted to the effect of a lateral confinement on the properties of granular flows. Both experimental and numerical studies (in two- and three-dimensional configurations) have pointed out that frictional lateral walls alter the flow properties~\citep{Azanza1999,Taberlet2003, Taberlet2004b, Jop_JFM_2005, Bi2005, Bi2006, Richard2008, Crassous_JSTAT_2008, Taberlet2008, deRyck2010, Brodu_PRE_2013, Brodu_JFM_2015, Artoni_PRL_2015, Yang_granularmatter_2016, CourrechduPont_PRL_2005, Holyoake_JFM_2012, Zhang_EPJE_2019,Richard_GM_2020,Zhu_GM_2020}. For example, steady and fully developed flows (SFD flows) have been observed up to large angles of inclination where accelerated ones are usually expected. These flows can be very different from usual dense Bagnoldian flows. They can have complex internal structures, including secondary flows and heterogeneous particle volume fraction.
Another interesting feature of laterally bounded flows over inclines is that, at any inclination angle, there is a critical flow rate above which the flow occurs on a static heap which forms along the base~\citep{Taberlet2003}. The angle of the heap, in a stationary state, is determined by the flow rate, the internal friction of the flow and its friction with the sidewalls. This transition to surface flow over the heap also induces strong alteration of the flow properties. Stationary flows atop this sidewall-stabilized heap (SSH) differ fundamentally from SFD flows on a bumpy base as they occur over erodible bases. 
It was shown~\citep{Taberlet2003} that these SSH are dynamically stabilized by the flow at their surface and that solely sidewall friction is responsible for their formation. Using a balance of momentum for the “flowing  layer” rubbing and sliding on the sidewalls, and assuming that both the solid volume fraction $\nu$ and the effective sidewall friction coefficient of the flowing layer are constant, one can derive the SSH equation, {i.e.} an approximate linear scaling law linking the free surface angle, $\theta$, the height of the flowing layer, $h$, and the width of the channel, $W$:

\begin{equation}
\tan\theta = \mu_{b,h} + \mu_{w,h} \frac{h}{W},\label{eqn:SSH}
\end{equation}
where $\mu_{b,h}$ is the effective internal friction coefficient of the flowing layer of depth $h$ on the heap and $\mu_{w,h}$ is the effective friction coefficient of the flowing layer on the sidewalls. 
At low angle $\theta$, near jamming, the above-mentioned assumptions are well verified and it is relatively easy to conceive a flowing layer whose height can be measured. For example Jop et al. ~\citep{Jop_JFM_2005} measured the flow thickness $h$ in their experiments, using a method based on erosion by the flow of a blackened metal blade. The SSH equation agrees with their experimental results. More surprisingly, 
it agrees also with experimental results obtained at much higher values of $\theta$ ~\citep{Taberlet2003}, for which the solid volume fraction is not
constant but varies strongly through the flowing layer to reach negligible values in the dilute region at the surface ~\citep{Delannay2007}. \revPR{It is possible to simulate such types of flows by means of Discrete Element Method simulations with periodic boundary condition along the main-flow direction (see figure~\ref{fig:resume}b) as showed in \citep{Taberlet2008}.} These  simulations~\citep{Richard2008} reveal a gradual weakening of friction at the sidewalls, thus the assumption of a constant effective friction on the wall is thus also questionable. Moreover, at high angles, the top of the flow is very dilute, with ballistic trajectories of the grains. At the bottom of the flow, the transition to rest is smooth. This makes difficult to define a height of the flowing layer. In~\cite{Taberlet2003}, the height of the flowing layer is determined using the quasi-linear velocity profile of the surface flow, but this definition is somewhat arbitrary. 
For large free surface angles, \cite{Richard2008} have shown that the solid volume fraction increases with depth
({i.e.}, from the free surface to the bottom of the system) with a characteristic length scale $l_\nu$ (see Fig.~\ref{fig:resume}c), and that the velocity scales linearly with the same length (see Fig.~\ref{fig:resume}d). 

These numerical simulations also revealed that the sidewall friction, which weakens with depth above a creeping region, also scales with $l_\nu$ (see Fig.~\ref{fig:resume}e). In the creeping region, beneath a critical depth which also scales with $l_\nu$, the velocity is weak and decreases exponentially with depth, the solid volume fraction has a constant value $\nu_0 \simeq 0.6$ and the sidewall friction is also constant, with a value significantly smaller than in the flowing zone. Using a balance of momentum for the region above this critical depth, it was possible to establish the validity of the SSH equation, using the length $l=2l_\nu$ as an effective height of the flow, for large free surface angles~\citep{Richard2008}: 
\begin{equation}
\tan\theta = \mu_{b,l} + \mu_{w,l} \frac{l}{W}\>, \label{eqn:SSHl}
\end{equation}
where the coefficients $\mu_{b,l}$ and $\mu_{w,l}$ are associated with the effective friction of the flowing layer of height $l$ with the creeping zone and the sidewalls, respectively.

Sidewall effects are not limited to small channel widths since they have been experimentally quantified for SFD flows on top of a static pile in wide channels (up to 600 particle diameters)~\citep{Jop_JFM_2005}. For wide channels, it is more difficult to obtain flows with large free surface angles, as it would require extremely large flow rates. In~\cite{Jop_JFM_2005}, all the results are obtained for values of $\theta$ smaller than $26^\circ$ except for one at $\theta = 33^\circ$. In low flow rate conditions, it is possible to give an asymptotic relation for the velocity $V$, derived from the $\mu(I)$ rheology~\citep{Jop_JFM_2005}: 
\begin{equation}
\frac{V}{\sqrt{gd}} \propto \left(\frac{W}{d} \right)^{3/2} \left(\frac{h}{W}\right)^{5/2},
\end{equation}

where $d$ is the diameter of the grains, with a good agreement with experimental measurements.
The scaling in $h$, the height of the flow, is different here from the linear one obtained in~\cite{Richard2008} by numerical simulations at large angles. This suggests  that $\mu(I)$ rheology is not valid for granular flows atop a SSH at large angle $\theta$.

The present article is dedicated to a numerical 
study of these confined SFD flows over an erodible bed, and more specifically to a better characterization of the scaling laws for the main characteristics of the flow. The scaling in $W$ has never been studied for large values of the angle $\theta$. This article is the first of a series. The second one will be dedicated to the characterization of the stress tensor and of rheology for confined granular flows over SSH.\\
The article is organized as follows. We present the simulation scheme in Sec.~\ref{sec:DEM}, detailing the interparticle force laws and the geometry of the studied system. In Sec.~\ref{sec:macro}, we report the discharge law of our system with varying interaction parameters (friction and restitution coefficients) and channel width. Section~\ref{sec:kine} is devoted to the kinetic properties of the studied granular flows including the behaviour of the volume fraction, velocity and granular temperature profiles for different channel widths. 
The transverse properties of the flows are investigated in Sec.~\ref{sec:transverse}.
Finally, in Sec.~\ref{sec:conclu} we summarize our findings and conclude.


\section{Methodology: the Discrete Element Method}\label{sec:DEM}

 The use of numerical simulations constitutes an interesting
alternative to experiments as they 
allow access to observables (stress tensors, individual trajectories and velocities\ldots) that cannot be measured easily.
Among the available methods, most used is the soft-sphere discrete element method (DEM). It has been widely used over the past decade to study granular systems in a wide range of geometries such as  flows down inclines ({e.g.}~\cite{Silbert2001,Richard2012}), surface flows ({e.g.}~\cite{Taberlet2008,Taberlet2004b}), shear cells ({e.g.}~\cite{Rycroft2009}), rotating drums ({e.g.}~\cite{Hill_PRE_2008,Taberlet2006a,Taberlet2006b,Richard2008b}) or silos ({e.g.}~\cite{Hirshfeld2001}).
\begin{figure}
\begin{center}
\includegraphics*[width=0.75\columnwidth]{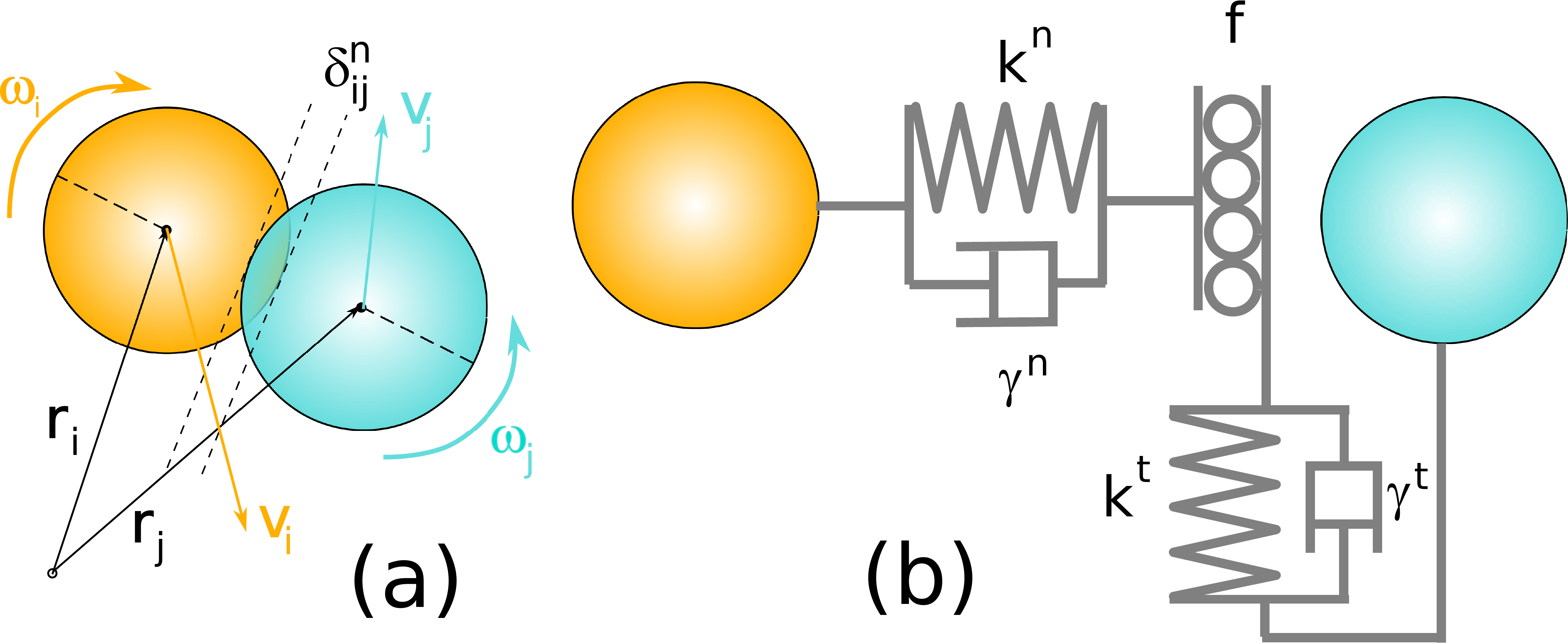}
\caption{(a) Two-dimensional sketch of two particles $i$ (radius $R_i$) and $j$ (radius $R_j$) at contact. The respective velocities are $\bf{v_i}$ and $\bf{v_j}$ (translational) and $\bf\omega_i$ and $\bf\omega_j$ (rotational).
The overlap between the two grains is denoted by $\delta_{ij}^n$.  (b) Sketch of the contact forces used. The normal force at the contact is modelled as viscoelastic (normal stiffness $k^n$, normal damping $\gamma^n$). The tangential force is also modelled as viscoelastic
(normal stiffness $k^n$, normal damping $\gamma^n$) but the elongation of the tangential spring is bounded
to a maximum value which is chosen according to a
Coulomb friction law (characterized by a friction coefficient $f$) when slip occurs.
}\label{fig:sketch_force}
\end{center}
\end{figure}
Here, each grain is a sphere  of radius $R_i$, mass $m_i$, moment of inertia $I_i$, position $\mathbf{r}_i$, velocity $\mathbf{v}_i$ and angular velocity ${\boldsymbol{\omega}}_i$ (see figure~\ref{fig:sketch_force}). 
\revPR{The normal forces used are given by the standard spring-dashpot interaction model: 
$\mathbf{F}^n_{ij}=k^n\delta_{ij}^n\mathbf{n}_{ij}-\gamma^n\mathbf{v}_{ij}^n,$
where $k^n$ is the spring constant, $\gamma^n$ the damping coefficient and $\mathbf{v}_{ij}^n$ the normal relative velocity.
Likewise we model the tangential force as a linear elastic and linear dissipative force in the tangential direction: $\mathbf{F}_{ij}^t = -k^t\boldsymbol{\delta^t_{ij}}-\gamma^t\mathbf{v}_{ij}^t,$
where $k^t$ is the tangential spring constant, $\delta^t_{ij}$ the tangential overlap, $\gamma^t$ the tangential damping and $\mathbf{v}_{ij}^t$ the tangential velocity at the contact point in the case of small overlap: 
$\mathbf{v}_{ij}^t=\mathbf{v}_{ij}-(\mathbf{v}_{ij}\mathbf{n}_{ij})\mathbf{n}_{ij}-
\left(R_i\boldsymbol{\omega}_{i} +  R_j\boldsymbol{\omega}_{j}\right)\times \boldsymbol{n_{ij}}.$
 The tangential overlap is set to zero at the initiation of a contact and its rate of change  is given by 
the tangential relative velocity.
We truncate the magnitude of $\delta_{ij}^t$ as necessary to satisfy Coulomb law {{i.e.}} $\left|\mathbf{F}_{ij}^t\right|\leq f_g\left|\mathbf{{F}}_{ij}^n \right|$, where $f_g$ is the grain-grain friction coefficient. Note that we assume that the static friction coefficient is equal to the dynamic one and that this friction coefficient depends neither on velocity nor on aging~\citep{Bureau2002}.
}
The
granular material is an assembly of $N$ dissipative spheres
(average diameter $d=2R$ and average mass $m$) submitted to gravity $\bf{g}$. A small polydispersity
of $\pm 20\%$ is considered to prevent crystallization. The following
values of the parameters are used:  $k^n d /mg = 5.6\times 10^6$, $k^t=2k^n/7$, $\gamma^t=0$. 
The value of $\gamma^n$ is adjusted to obtain the desired value of the normal restitution coefficient $e^n$. 
All the simulations are carried out with $e^n=0.88$ and $f_g=0.5$.\\
The simulation chute  (see figure~\ref{fig:resume}b) consists of a three-dimensional cell
which can be inclined relative to the horizontal by an angle $\theta$ (angle between the horizontal and the main-flow direction : $x$).
Its size  in the $x-$  direction is set to $L_X=25.3d$ with periodic boundary conditions in this direction. In the $z-$ direction ({i.e.} normal to the free surface of the flow)
 the size of the cell is set to large values and can be considered as infinite.
Frictional sidewalls are located at positions $y=-W/2$ and $y=W/2$. They are treated as sphere of infinite radius and mass. The restitution coefficient is the same as that used for the grain-grain interactions. The friction coefficient between grains and sidewall $f_w$ is set to $0.5$. 
The  bottom of the cell is made bumpy by pouring under gravity $\mathbf{g}$ a large number of grains into the cell and by gluing those that are in contact with the plane $z=0$. \\
\revPR{
Since the flows occur on a quasi-static pile, it is important to explicit the initial conditions which, potentially, have an important influence on the properties of the creeping zone. We chose to built our system from an energetic flow which slow down progressively until it reaches a steady state. 
Thus the creeping zone originates from an accretion process similarly to what is done in \cite{Taberlet2003}.
Our procedure is the following. 
We first set the angle of the system to a large value ($\theta=70^\circ$) and the grains are initially arranged in the simulation cell in an ordered slightly dilute ``hexagonal compact packing''.  
Each components of the grains' velocity is initially randomly assigned. The distribution used is uniform between $\sqrt{2g R}$ and $-\sqrt{2gR}$. This leads after a short time  (approximately $100\sqrt{d/g}$) 
 to an accelerated energetic system without signs of the initial ordered structure which flows over its entire depth.
The angle was then decreased instantaneously to the desired value.
After a long  transient (several hundreds of $\sqrt{d/g}$), a SSH ({i.e.} a SFD flow on a quasi-static layer)  is obtained.
Note that, the SFD-criterion corresponds to a stabilization of the total kinetic energy within fluctuations. The kinetic energy of the creeping zone being very low, this criterion is not sensitive to structural modifications of the latter zone whose evolution characteristic time is very large with respect to the inverse of the shear rate of the flowing zone.
%
As mentioned above, due to the presence of a quasi-static zone,  the way the system is built has probably an influence on the properties of the system. 
We have checked that other protocols ({e.g.} use of an initial horizontal system with frictionless particles to obtain a static initial packing at random close packing)  lead to significantly different results
in the creeping zone (fabric, stresses in particular) but seem to have a moderate influence on the flowing zone.
}\\
Our results are given in non-dimensional quantities by defining
the following normalization parameters: distances,
times, velocities, forces, elastic constants and viscoelastic constants are,
respectively, measured in units of $d$, $\sqrt{d/g}$, $\sqrt{gd}$, $mg$,  $mg/d$ and $m\sqrt{g/d}$.\\
In the remainder of the paper several quantities will be reported and studied by means of profiles of kinematic quantities.
The variables obtained from the DEM simulations ({e.g.} grain velocities) are averaged to obtain continuum quantities. 
Given a grain quantity $\psi_p(\mathbf{r_p},t)$ its average $\langle\psi\rangle$ over a given spatial region $\mathcal{V}$ of volume $V$ is obtained both by space and time averaging:
\begin{equation}
\rho \langle \psi \rangle = \frac1V
\frac{1}{N_t} \sum_{N_t} \sum_{r_p\in\mathcal{V}} m_p \psi_p.
\end{equation}
The fluctuations ({e.g.} granular temperature) are defined by subtracting the effect of the local gradient of the studied quantity as shown in \cite{Glasser_PoF_2001,Xu2004} and \cite{Artoni_PRE_2015}. 
In the remainder of the paper, two types of profiles are reported in the case of steady ({i.e.} time independent) and fully developed ({i.e.} space independent in the main flow direction) flows.
First, profiles for which the studied quantity is averaged over the $x-$ and $y-$ directions ({e.g.} $v_x(z)$ for the evolution of the streamwise velocity versus the vertical position $z$). They are obtained by
averaging it over time and over space using parallelepiped volumes (height $d$, width $W$ and length $L_X$)
centred at $x=L_X/2$, $y=W/2$ and $z$. Second, profiles averaged only over the $x-$direction ({e.g.} $v_x(y,z)$)  are obtained similarly by averaging the studied quantity over time and space using parallelepiped volumes (height $d$, width $d$ and length $L_X$)
centred at $x=L_X/2$, $y$ and $z$.  
\section{Macroscopic description of the granular flow}\label{sec:macro}
In this section we will mainly focus on the output flow rate {i.e.} a macroscopic property defined globally at the scale of the system. By contrast, section~\ref{sec:kine} will be devoted to properties of the flows at the local scale.
\subsection{Steady fully developed flows and Phase diagram}\label{sec:diag}
We consider steady flows over an erodible bed, {i.e.}, flows whose properties are independent of time regardless of the fluctuations.
To achieve such a flow, in addition to frictional sidewalls, a sufficient number of grains has to be used to obtain a flow rate greater than the critical one reported in~\citep{Taberlet2003}.  
\begin{figure}
\begin{center}
\resizebox{0.75\columnwidth}{!}{\includegraphics*{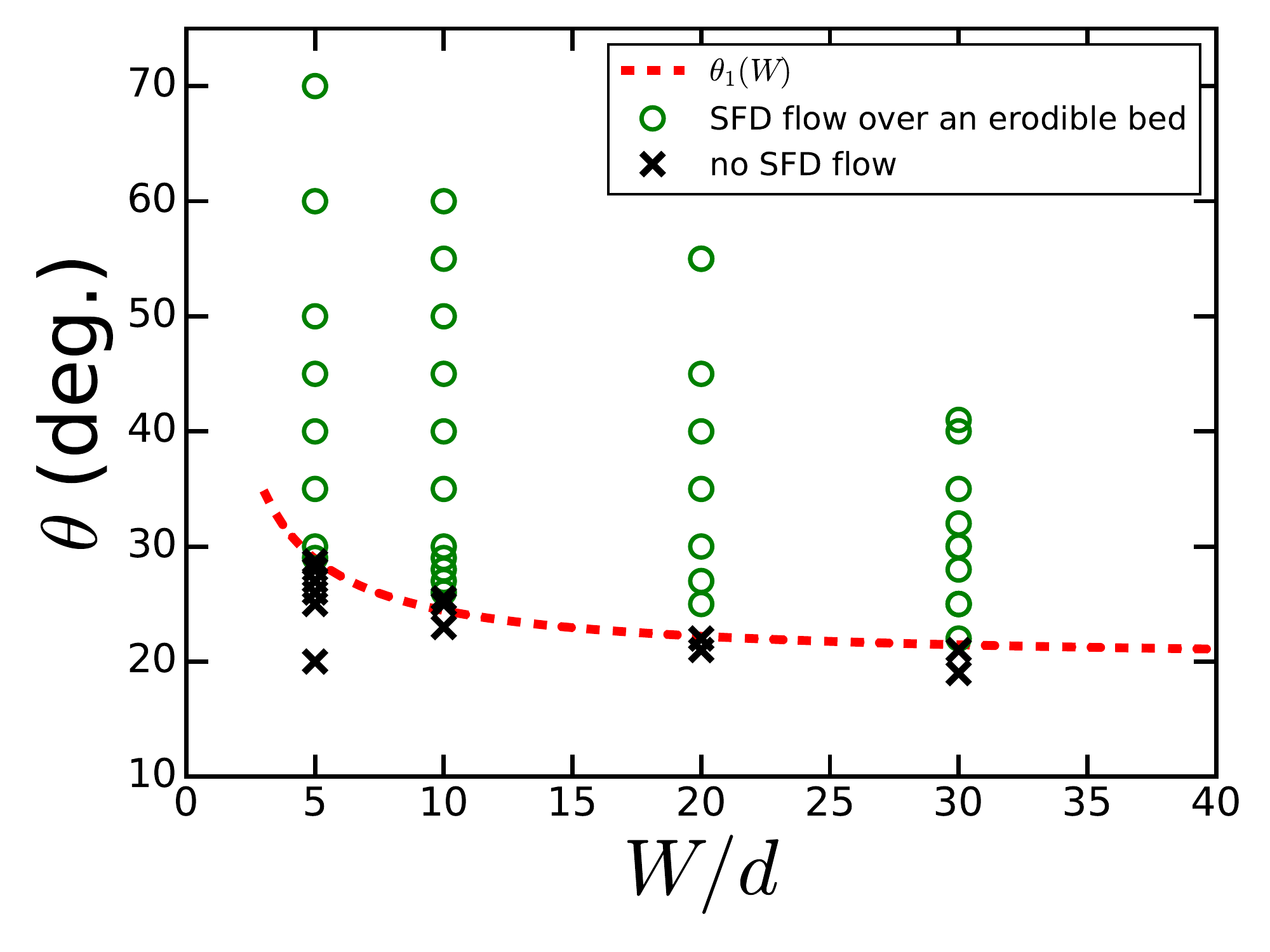}}
\caption{For a given gap between the sidewalls, a SFD flow over an erodible bed can only be observed above a given inclination angle. Below the latter, the system can experience an intermittent surface flow or be at rest. This is illustrated in the phase diagram reported in the parameter space $(W/d,\theta)$. 
The dashed line represents the angle $\theta_1(W)$ above which a SFD flow can be achieved.}
\label{fig:phase_diagram}       
\end{center}
\end{figure}
At $t=0$ 
the angle $\theta$ is instantaneously decreased from $70^\circ$ to
the desired value. 
After a transient corresponding to a global slow down of the system, 
the number of grains flowing out of the cell per unit time, $\partial_t N_{out}$, reaches a plateau, indicated in the figure by a dashed horizontal line. We have verified that this plateau corresponds to the occurrence of a SFD flow 
throughout the simulation cell. The value of the flow rate obtained at  the plateau corresponds to the SFD flow rate.
We have checked that this flow rate does not depend on the number of grains as long as the latter is large enough to obtain an erodible bed~\citep{Richard_AIPConf_2010}. 
In this case, if the number of grains is increased, the erodible bed becomes thicker but the flow zone and thus the flow rate do not change.
In our simulations the creeping region corresponds to at least 10 grain layers. 
\\ 
Within the control parameters investigated so far ($20^\circ \leq\theta \leq 70^\circ$
and $5\leq W/d\leq 30$), we can obtain different regimes. If the angle is large enough, a SFD flow is reached after a transient, while at low angles the system remains completely jammed, {i.e.}, without flow. 
This is illustrated in figure~\ref{fig:phase_diagram}, where the phase diagram is reported. 
\revPR{In addition, we have indicated the lower limit of the existence of steady flows : $\theta_1(W)$.}
Note also that, in vicinity of the latter, the temporal evolution of the flow rate may be intermittent, leading to a significant uncertainty in the determination of $\theta_1$. Indeed, an intermittent behaviour may be observed for a long period of time before the system comes to rest. 
The study of this intermittency is outside of the scope of this paper and would require an intensive statistical analysis, as done in \cite{Lemieux2000}. 
It is also important to emphasize here that the presence of flat frictional walls allows for the existence of SFD flows
at much higher angles than those observed for unconfined flows~\citep{Silbert2001}. Besides, even at very steep angles the flow reaches a
steady state~\citep{Taberlet2003,Brodu_PRE_2013, Brodu_JFM_2015}. 
These results point out the crucial importance of sidewalls in the behaviour of granular flows.
Additionally, as already been reported in the literature~\citep{Liu1991,Courrech2003,Metayer2010}, one can note that the lower limit angle $\theta_1$ for steady states depends on the lateral width $W$:
the wider the gap, the smaller $\theta_1$. 
This variation is the most important for $W<20d$ due to sidewall effects.
\subsection{Flow rate $Q$ versus $\theta$ and $W$}\label{sec:Q_vs_theta_W}
\begin{figure}
\begin{center}
\resizebox{0.85\columnwidth}{!}{\includegraphics*{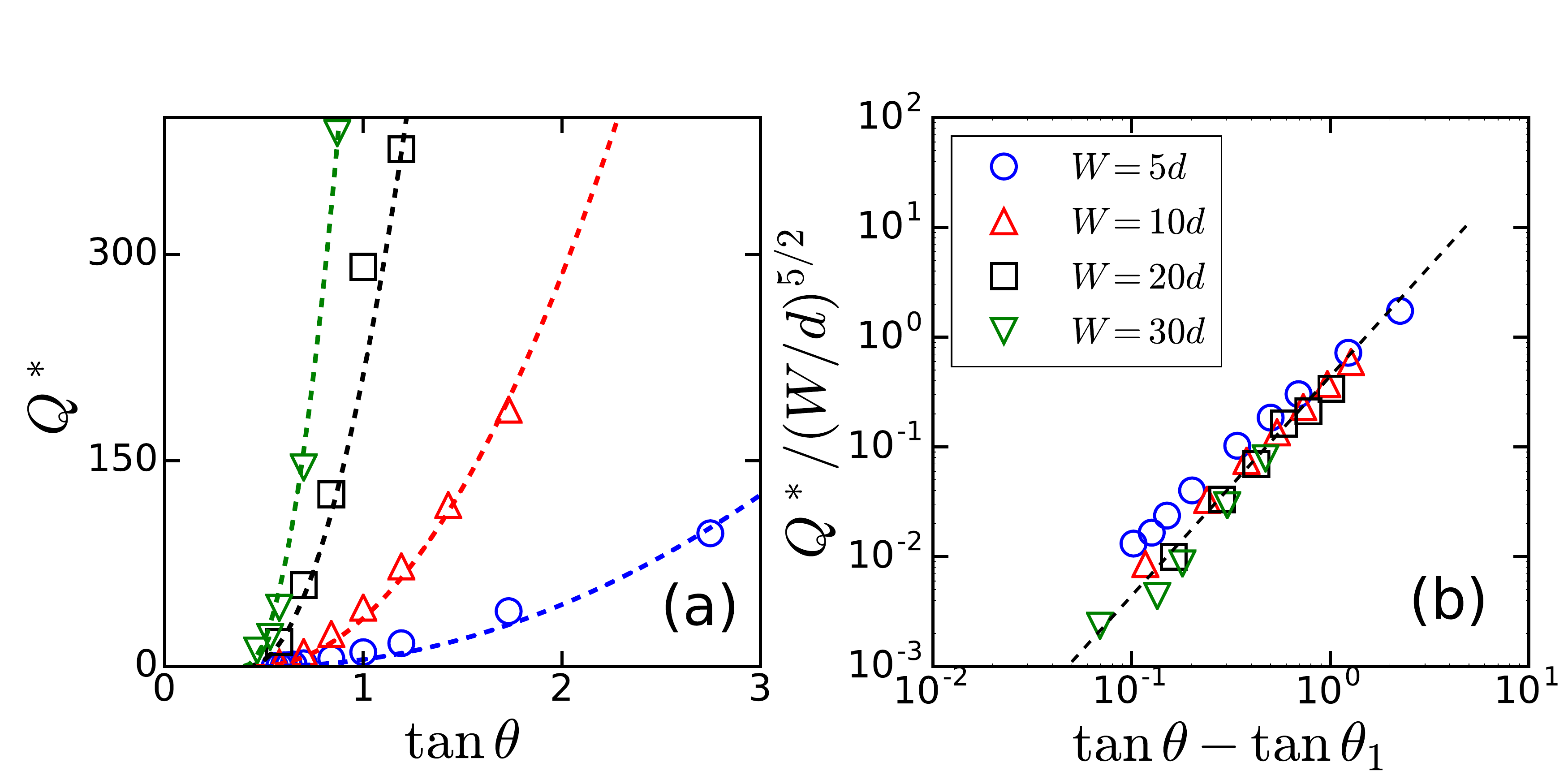}}
\caption{(a) For any gap between sidewalls $W$, the flow rate per unit width, $\Qadim$, is found to increase with the angle of the free surface $\theta$. For a given $\theta$, increasing $W$ leads to an increase of $\Qadim$. (b) The flow rate per unit width can be rescaled by $(W/d)^{5/2}$. The evolution of the rescaled flow rate versus $\tan\,\theta$ can be fitted by a quadratic law. 
}
\label{fig:Q_vs_theta}       
\end{center}
\end{figure}

The flow rate per unit width, $Q$, is an important macroscopic quantity characterizing the ability of the system to flow. 
It is defined as $Q=\Delta m/(\Delta T W \rhop)$, where $\Delta m$ is the mass flowing out of the cell during the time $\Delta T$ and $\rhop$ the grain density.
We have plotted in figure~\ref{fig:Q_vs_theta}.a
the dimensionless flow rate per unit width, $\Qadim = Q / d\sqrt{gd}$, versus
the tangent of the angle of inclination $\theta$ for different gap widths $W$. 

For a given angle of inclination, 
the flow rate per unit width $\Qadim$ is an increasing function of the gap width $W$ 
as a consequence of 
the declining influence of the wall friction~\citep{Taberlet2003,Jop_JFM_2005}.
For a given gap width, a  one-to-one  relation, which increases monotonically,  exists between the flow rate and the angle.
It is a consequence of sidewall friction and the increase of $\theta$ with $\Qadim$  is more pronounced for low values of $W$
and  is expected to be very weak for very large channels~\citep{Jop_JFM_2005}.
Our results are compatible with experimental ones~\citep{Khakhar2001,Courrech2003,Taberlet2003,Jop_JFM_2005,Metayer2010} that report an increase of the angle of the flow with the flow rate. 
One can note again
that the lower limit angle $\theta_1$ below which the flow stops is a decreasing function of the gap width (see section~\ref{sec:diag}).
The numerical evolution of $\Qadim$ with $W$ is compatible with a $W^{5/2}$-scaling (figure~\ref{fig:Q_vs_theta}(b)) consistent with that obtained by \cite{Jop_JFM_2005} for much lower angles and larger gap widths.
Also, its evolution with $\tan\theta$ can be fitted by a parabolic law $Q\propto (\tan\theta-\tan\theta_1)^2$.

\section{Kinematic properties of the flow}\label{sec:kine}


\subsection{Volume fraction profiles and  definition of the flowing height}\label{sec:compa}

\begin{figure}
\begin{center}
\includegraphics*[width=0.95\columnwidth]{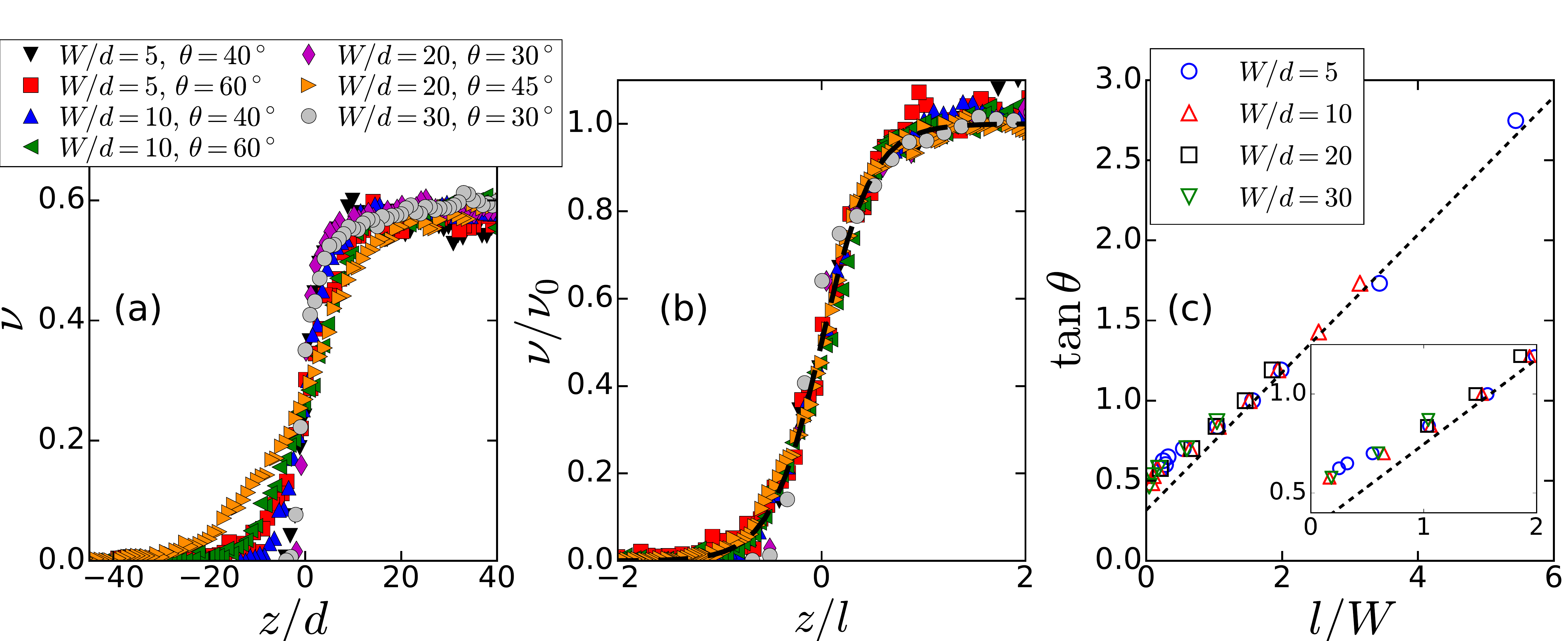}
\caption{The vertical profiles of the volume fraction for various inclination angles $\theta$ and widths $W$ (a) collapse onto a  master curve 
$\nu(z) = \frac{\nu_0}{2}\left[  1 +\tanh \left( {2z}/{l}\right)  \right]$ --dashed line in panel (b).
The characteristic length $l$ scales with $W$ and increases with the angle of inclination $\theta$ (c). The dashed line represents the best linear fit for all the data. At low angle a slight deviation from the linear relation is found (inset of panel (c)). The angle for which this deviation is visible is close to $\theta_c=40^\circ$ and $l/W\approx 1$. For this reason, the linear fit reported is done for angles above $\theta=40^\circ$. 
}
\label{fig:profil_compa}       
\end{center}
\end{figure}

Unidirectional dense granular flows down bumpy inclines are characterized by an almost constant volume fraction 
throughout the flow height~\citep{Silbert2001,Bi2005,Bi2006,Delannay2007,Brodu_PRE_2013}.
Figure~\ref{fig:profil_compa}.a shows vertical profiles of the volume fraction, $\nu$, averaged across sidewalls
for various inclination angles ($30^\circ \le \theta \le 60^\circ$) and lateral widths ($5d \le W \le 30d$). 
Contrary to dense flows sheared over the whole depth (hereafter referred as fully mobilized flows)  an erodible bed 
forms at the bottom of the system and, consequently, 
the volume fraction is, in most cases, far from being constant and varies from $0$ to approximately $0.6$.  
Only for low flow angles ({e.g.} $\theta=30^\circ$ for $W/d=30$)  do the volume fraction profiles tend toward  a step function that is a constant volume fraction.\\

 
Remarkably, the vertical profiles of the volume fraction, $\nu$, for different inclinations and widths collapse onto a single curve
(see figure~\ref{fig:profil_compa}.b) \revPR{which can be fitted by the following equation~\citep{Richard2008}:}
\beq
\nu(z)=\frac{\nu_0}{2} \left[  1 +\tanh \left( \frac{2z}{l}\right)  \right] ,
\label{nu}
\eeq
with the origin such that $\nu = \nu_0/2$ at $z=0$. The characteristic length of variation of the volume fraction $l$ depends on both the gap width and flow angle and, in most cases, it is greater than the grain size. \revPRb{It should also be pointed out that the validity of equation~\ref{nu} is questionable at low values of the characteristic length $l$ due to the possible intermittency of the flow.} \revPRb{The length $l$ scales with $W$ and increases with inclination and equation~(\ref{eqn:SSHl}) can be rewritten as:}
\beq
\frac{l}{W}=\frac{1}{\mu_{w,l}}\left(\tan\theta - \mu_{b,l}\right).
\label{eqn:h_theta}
\eeq
We have verified this equation for  $W/d = 5,10,20$ and $30$ and obtained 
$\mu_{b,l} = 0.32$
and $\mu_{w,l}= 0.43$
(see figure~\ref{fig:profil_compa}.c).  
Note that the characteristic length $l$ used here is twice the length $l_\nu$ used by~\cite{Richard2008}. 

The variation of $l/W$ with $\tan \theta$ deviates from an affine law for $\theta < \theta_c \approx 40^\circ$; note that $\theta_c$ is the angle for which $l/W\approx 1$.
As will be shown below, $\theta_c$ corresponds to the boundary between a dense flow regime (regime I) and a regime for which a strong variation of the volume fraction is observed within the flow (regime II). It is not easy to say whether the value of $\theta_c $ depends on $W$. There is likely a dependency, 
but it seems rather weak, with a small increase of $\theta_c $ (only a few degrees) when $W$ decreases.
A similar but more pronounced effect was observed for $\theta_1$ (see figure~\ref{fig:phase_diagram}).

\subsection{Contact number}\label{appA}
\begin{figure}
\begin{center}
\resizebox{0.95\columnwidth}{!}{\includegraphics*{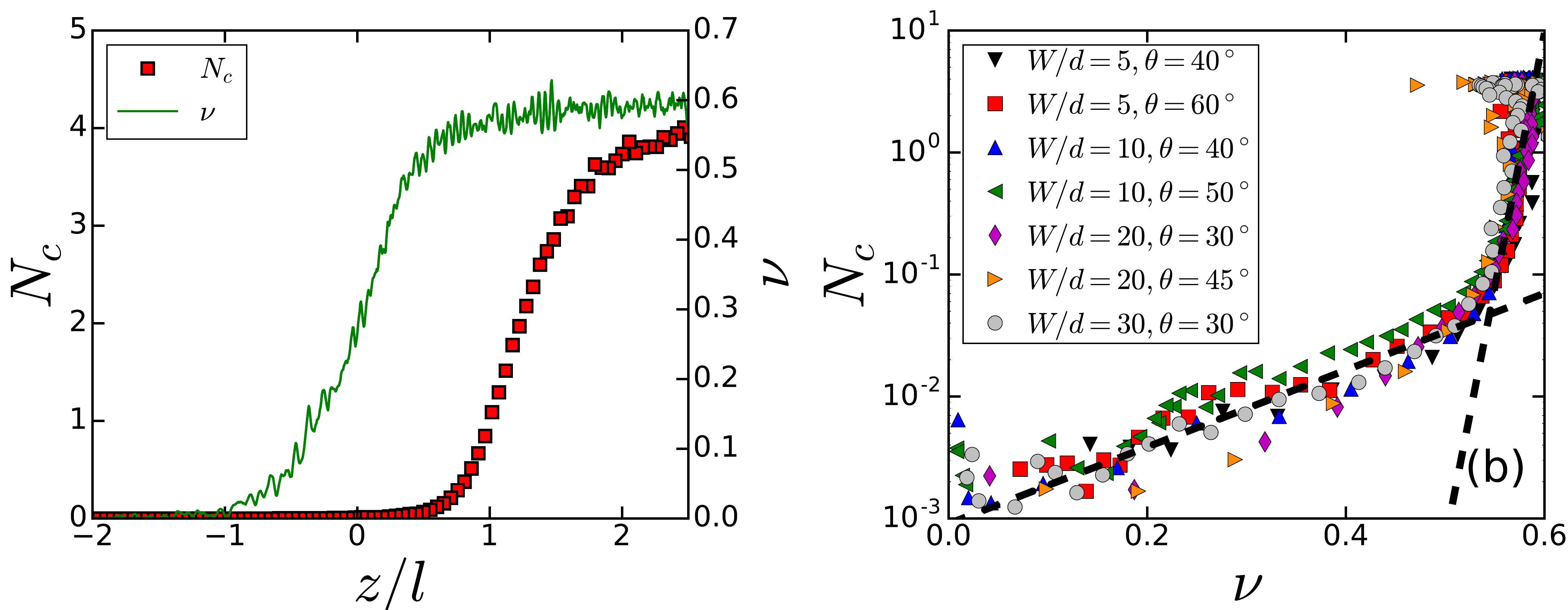}}
\caption{(a) Variations of the number of contacts, $N_c$, and of the volume fraction $\nu$ with the flow depth. (b) The gap between sidewalls is $W=10d$ and the angle of the flow $\theta=50^\circ$. The grain-grain and grain-wall friction coefficients are both equal to $0.5$. The variation of $N_c$ versus $\nu$ displays a universal behaviour with a kink corresponding to $\nu\approx 0.54$ and $N_c\approx 0.06$ (b).
}
\label{fig:contacts}       
\end{center}
\end{figure}

Another interesting property of  a granular flow is the average number of contacts per grain, $N_c$. We have illustrated its variation throughout the
flow and its link with the volume fraction in Figs.~\ref{fig:contacts}(a) and (b), respectively. 
As expected, it increases with increasing depth from zero in the gaseous zone
to $N_c\sim 4$ within the creeping zone, which corresponds to  the number of contacts obtained for an iso-static packing {i.e.} a packing for which all the contacts are perfectly frictional. 
We have explained in section~\ref{sec:DEM} that the initial state of the system might influence the reached steady state. The protocol used in the present work indeed leads to relatively loose (and fragile~\citep{Berzi_SM_2019}) systems but others might possibly lead to denser systems and thus to a number of contacts larger than $4$ in the creeping zone.
%
It should be pointed out that these quantities ($N_c$ and $\nu$) vary with the same length scale $l$ but they are shifted vertically: in the example reported in figure~\ref{fig:contacts}
the number of contacts reaches the half of its maximal value at $z/d\approx 22$ (depth for which $\nu$ tends toward $0.6$)
whereas $\nu=\nu_0/2$ at $z/d=0$ (depth for which $N_c$ is below $0.1$).
Thus, in the flowing zone ($z\in[-l/2,l/2]$), the flow is mainly collisional. 
When we plot the number of contacts as a function of the volume
fraction, two different exponential regimes can be clearly identified: one for $\nu\lesssim 0.53$ for which the average number of contacts is very small (below $0.1$) and the other one for $\nu\gtrsim 0.53$ where the average number of contacts increases rapidly with $\nu$. 
As $N_c$ and $\nu$ vary over the same length scale $l$, the relationship between these quantities is independent of $\theta$ and $W$: the regime change corresponds to the transition between the buffer zone and the flowing zone which occurs at $z=l/2$, {i.e.}, at $\nu(z=l/2)=\nu_0/2\left[ 1 + \tanh (1)\right]\approx0.528$. 
Note also that this curve could depend on 
the grain stiffness $k^n$ used to model the normal force between grains (see Sec.~\ref{sec:DEM}). It has indeed been shown that this quantity plays an important role on the number of contacts within granular flows, which increases with decreasing stiffness~\citep{Ji2008,Vescovi_SM_2016,Vescovi_GM_2018}. The volume fraction can also be influenced by the grain stiffness~\citep{Vescovi_SM_2016} especially in dense zones for which the overlap between contacting grains might have a non-negligible volume. A full study of the effect of grain stiffness on our results is outside the scope of the paper. Yet, to position our study with respect to that of~\cite{Vescovi_SM_2016} we have estimated  the dimensionless parameter $K^*=k^n/\rhop d^3 \dot\gamma$, $\rho_p$ being the grain density. Since $\dot \gamma$ varies within our system, so does $K^*$. Yet the order of magnitude of its lowest value is $10^6$. For  such a value, \cite{Vescovi_SM_2016} have found an effect of the grains' stiffness on the volume fraction only if the latter exceed $0.625$ which is never the case in our simulations.

\subsection{Velocity profiles}\label{sec:vel}
\begin{figure}
\begin{center}
\resizebox{0.95\columnwidth}{!}{\includegraphics*{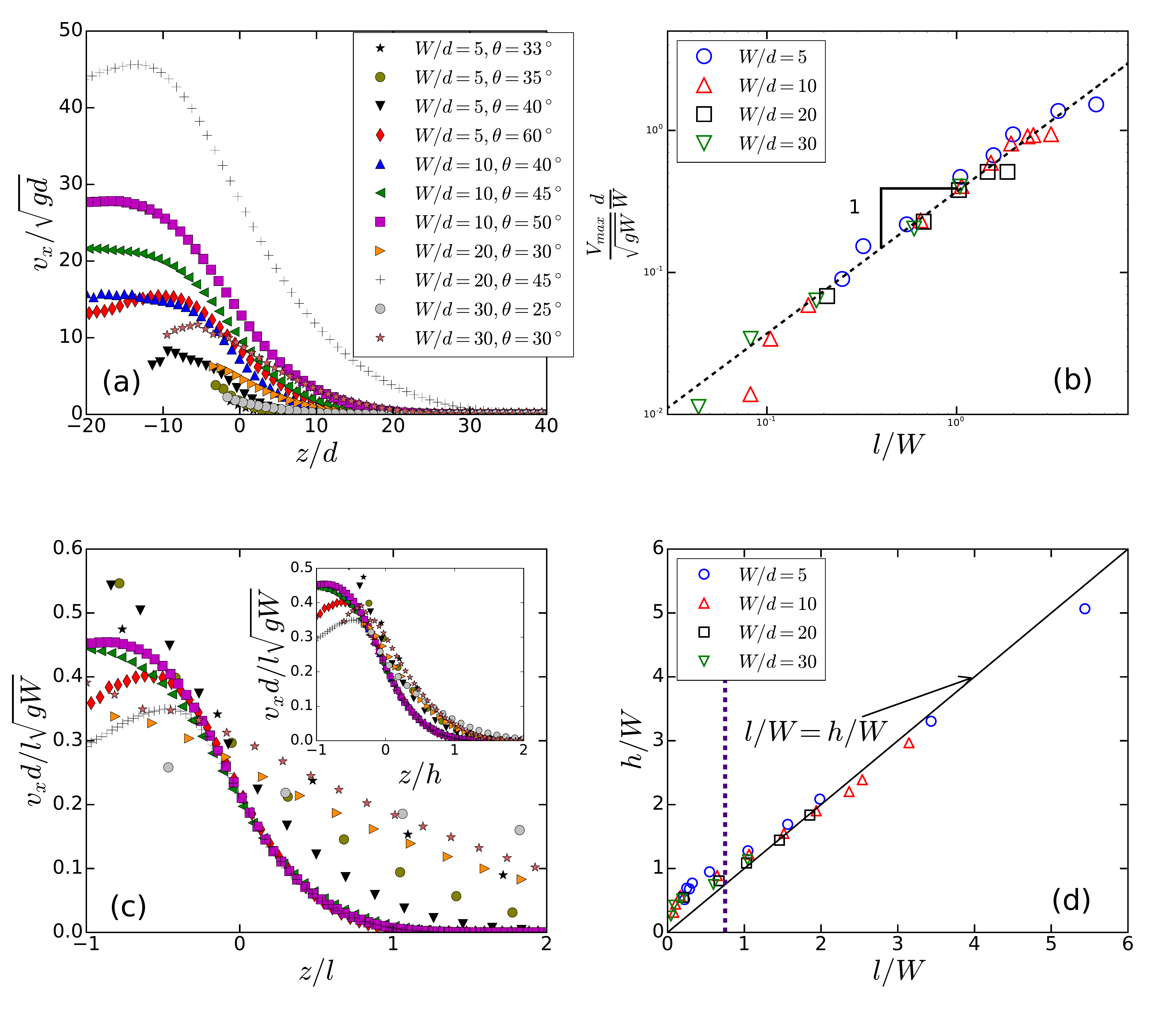}}
\caption{(a) Vertical profiles of the  particle velocity  in the main flow direction for various inclination angles $\theta$ and widths $W$. 
(b) The maximal velocity (estimated by $v_x(z=l/2)$) scales with the characteristic velocity
$\sqrt{gW}(l/d)$.
(c) Flows in regime II collapse onto a single curve when rescaling the velocity by $\sqrt{gW}l/d$ and the depth by $l$ or by $h$ (inset of panel(c)).
(d) In regime II, the characteristic length $l$ determined by fitting the volume fraction profiles  is close to  $h$, the length determined through the velocity profile. In regime I, below $\theta_c \approx 40^\circ$ important relative differences are observed (c). 
}
\label{fig:profil_vit}       
\end{center}
\end{figure}

The vertical profiles of the streamwise velocity are reported in figure~\ref{fig:profil_vit}a for various gaps $W$ between the sidewalls  and various flow angles $\theta$. As expected, the two latter parameters have a strong influence: the velocities are found to increase with increasing angle and/or gap. Interestingly, at large angles, the velocity profiles display an inflection point close to $z/d=0$, around which the profile is reasonably linear.
\revPR{In the vicinity of the free surface, two cases are 
observed: either a strict maximum or a plateau. 
For a given $W$ we observe the plateau at low angles. When the angle is larger that a given value (which decreases strongly with increasing $W$) a strict maximum is observed. In the remainder of the paper, depending on the case, we will call $V_{max}$ the velocity of the strict maximum  or of the plateau.
}

\begin{figure}
\begin{center}
\resizebox{0.75\columnwidth}{!}{\includegraphics*{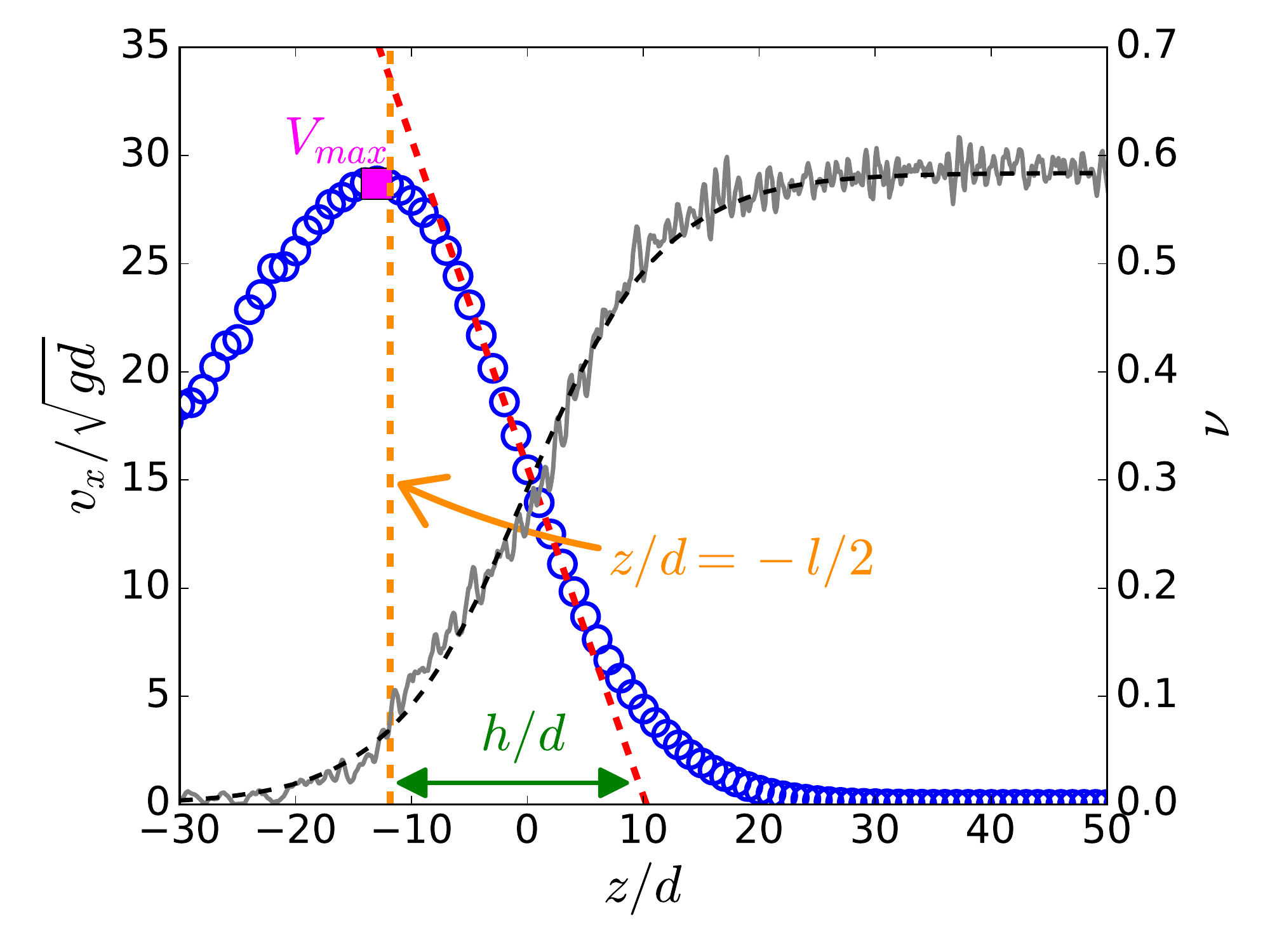}}
\caption{
Velocity (blue open circles) and volume fraction profiles (grey line) and its fit by equation~\ref{nu} (black dashed line) for
$W/d=10$ and $\theta=53^\circ$.
The height of the flow $h$ is the distance between $z=-l/2$ (orange dashed vertical line, position corresponding approximately to $V_{max}$) and the depth at which the tangent to the linear part of the velocity profile (dashed red line) intersects the $z$ axis.
}
\label{fig:sketch_mesure_h}       
\end{center}
\end{figure}

Following ~\cite{Richard2008}, we rescale the velocities and the vertical positions respectively by $\sqrt{gW}(l/d)$ and $l$ (figure~\ref{fig:profil_vit}c). The scaling holds for angles larger than $\theta_c$ (regime II). Interestingly, the maximum velocity $V_{max}$, which corresponds approximately to $V(z=-l/2)$, is found to scale always (in regime I as in regime II) with the characteristic velocity $\sqrt{gW}(l/d)$ (figure~\ref{fig:profil_vit}b) which, consequently, can be considered as a characteristic velocity of the system. 
Figure~\ref{fig:profil_vit}c also points out that $l$ is not the right characteristic length for the variation of velocity with depth in regime I. We have thus to introduce a new characteristic length. \revPR{\cite{Taberlet2003} defined --somewhat arbitrarily-- the flow  height as twice distance between $z=0$ ({i.e.} the  position of the inflexion point) and the depth at which the tangent to the linear part of the velocity profile intersects the vertical axis ($z$ axis).} Taking into account the fact that the velocity profile is not symmetric according to $z = 0$ in regime I (see also figure~\ref{fig:flux}) and the characteristic behaviour of $V_{max}$, we define here the flow height $h$ as the distance between $-l/2$ (position corresponding approximately to $V_{max}$) and the depth at which the tangent to the linear part of the velocity profile intersects the $z$ axis (see figure~\ref{fig:sketch_mesure_h}). Rescaling the depth by $h$ (inset of figure~\ref{fig:profil_vit}c), leads to profiles in regime I which depend weakly on the inclination angle $\theta$,  as predicted by the $\mu(I)$ rheology \citep{Jop_JFM_2005}, and are not on the master curve of regime II which is independent of $\theta$.


Note that the flow obtained for $W/d=10$ and $\theta=40^\circ$ clearly belongs to regime II. In contrast, that obtained for $W/d=5$ and $\theta=40^\circ$ seems to belong to the end of regime I, it corresponds probably to a transition between regimes I and II. This suggests a slight dependence of the boundary between the two regimes \revPR{(and thus of the angle $\theta_c$)} on the gap between the sidewalls. 

As mentioned before, a peak in the velocity profile is observed for large angles and it should be pointed out that the scaling does not hold post-peak ({i.e.}, for $z$ lower than the peak's position).
This is the case, for instance, for ($\theta=60^\circ,\ W/d=10$) and for ($\theta=45^\circ,\ W/d=20$).
At  the top of the flow, we observe that grains are  mainly  submitted to collisions with sidewalls due to the very low volume fraction. The latter being immobile, the corresponding grains are slowed down in a more important manner than the grains located in the bulk which collide mainly with grains that move in the main flow direction. Also the grains located at the top of the flow slow down the grains located directly below them that, in return, also slow down the grains located below them. This global slow down ceases when the probability for a grain to collide with sidewalls is not significant. Thus, the behaviour of the granular flow in the vicinity of its top is complex and strongly influenced by the gap between sidewalls.

Figure~\ref{fig:profil_vit}d compares the two characteristic lengths of the flow : the characteristic length of the velocity profile $h$, and the characteristic length of the volume fraction profile $l$. At large angles, in regime II, $h/W$ and $l/W$ look identical. In regime I, below $\theta_c \approx 40^\circ$, $h/W$ becomes clearly larger than $l/W$, but the two lengths remain correlated. 
Thus, in regime I, $l$ does not correspond anymore to the flowing friction height.

\begin{figure}
\begin{center}
\resizebox{0.95\columnwidth}{!}{\includegraphics*{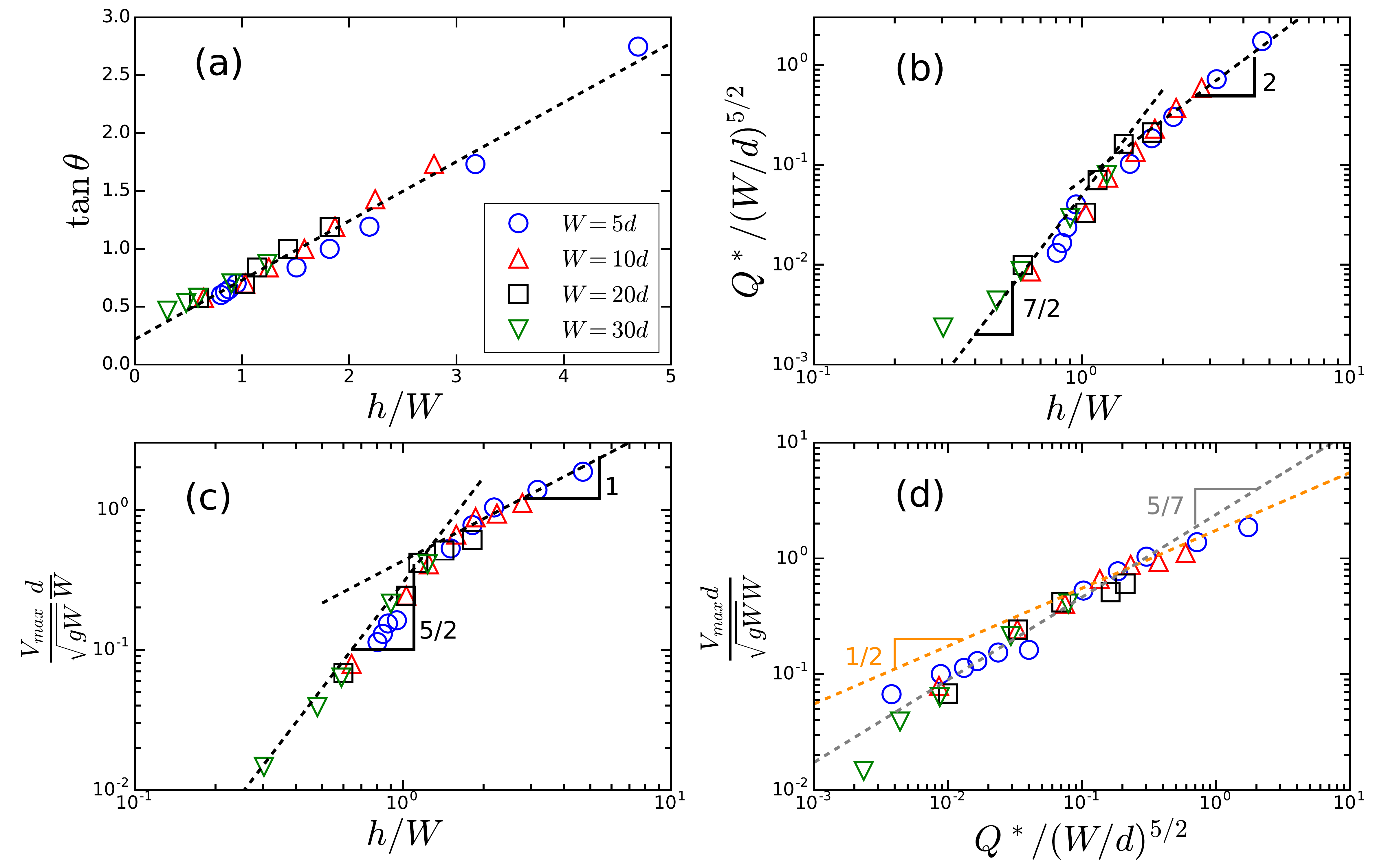}}
\caption{(a) The tangent of the angle of the flow depends linearly on the ratio $h/W,$ where $W$ is the gap between the sidewalls and $h$ the depth of the flow determined by fitting the velocity profile (see text). (b) The dimensionless flow rate per unit width $\Qadim$ scales with $(W/d)^{5/2}$ at any angle. Its dependence with $h/W$ is more complex: it is compatible with a $(h/W)^{7/2}$ scaling for angles below $\theta_c\approx 40^\circ$ (regime I) and a
$(h/W)^{2}$ scaling above (regime II). (c) Similarly, the maximal value of the velocity $V_{max}$ scales with $(W/d)^{3/2}$ at any angle and with $(h/W)^{5/2}$ ($\theta < \theta_c$, regime I) and $h/W$ ($\theta>\theta_c$, regime II). 
(d) \revPRb{The difference of scaling for the two regimes can also be confirmed by} plotting $V_{max} d / (W\sqrt{gW})$ versus $\Qadim/(W/d)^{5/2}$: the former quantity scales with the latter to the power $5/7$ and $1/2$ respectively for $\theta<\theta_c$ and $\theta>\theta_c$. 
}
\label{fig:tan_Q_Vmax_vs_hsurW}       
\end{center}
\end{figure}

Since $h$ characterizes the friction height for both regime I and regime II, we have reported in figure~\ref{fig:tan_Q_Vmax_vs_hsurW}a the evolution of the tangent of the flow angle versus the ratio $h/W$. The SSH equation (equation~(\ref{eqn:SSH})) is rather well verified. 
In regime I, the values of $l/W$ and $h/W$ are significantly different. Consequently, that of $\mu_{b,h}$ is different from that of $\mu_{b,l}$ (respectively $0.22$ and $0.32$)
and that of $\mu_{w,h}$ from that of $\mu_{w,l}$ (respectively $0.51$ and $0.43$).

Interestingly, when $l/W\to 0$, $h/W\to (h/W)_{min} > 0$. For a given $W$ there is thus a finite minimum value of $h$: $h_{min}(W)$, and this minimum value increases proportionally to $W$: 
$h_{min}(W) = W (h/W)_{min}$.
\revPRb{Note that
the limit $l/W \to 0 $ corresponds in our case to unconfined geometries ($W\to +\infty$), for which  
SSH flows have never been observed. 
The equation $h_{min}(W) = W (h/W)_{min}$ is consistent with the latter observation since $W\to \infty$ leads to $h_{min}\to \infty$. In other words, in unconfined geometries, SSH flows are only observed for infinitely high systems.} When $l/W\to 0$, the SSH equations (\ref{eqn:SSH}) and (\ref{eqn:SSHl}) 
lead to the relation: $\mu_{b,l}=\mu_{b,h}+\mu_{b,w} (h/W)_{min}$, from which we obtain $(h/W)_{min}=0.33$. 

It should be pointed out that $\tan\theta_1 > \mu_{b,l} > \mu_{b,h}$. 
All these quantities are probably close to each other when $W \to \infty$.
Note that $\theta_1$ is obtained directly from  simulations in which the system transits from a dynamical state to a static one, whereas the two others are obtained from  fits of numerical data obtained from the simulation of dynamical states. In the literature, $\mu_{b,h}$ is classically assumed to be gap width independent~\citep{Liu1991,Boltenhagen1999,Courrech2003,Metayer2010}. Yet, thanks to our extensive simulations, which provide results for wide ranges of both gap widths and flow angles, a small but systematic increase of $\mu_{b,h}$ with $W$ is found. In contrast, the angle $\theta_1$ decreases with increasing $W$ (see the phase diagram reported in figure~\ref{fig:phase_diagram}). This difference points out that these two quantities do not quantify the same physical mechanisms.\\ 

Following \cite{Jop_JFM_2005} we have studied the scaling of both $\Qadim$ (figure~\ref{fig:tan_Q_Vmax_vs_hsurW}b) 
and $V_{max}$ (figure~\ref{fig:tan_Q_Vmax_vs_hsurW}c) versus the gap $W$ between the sidewalls  and the height $h$ of the flowing layer.  In agreement with~\cite{Jop_JFM_2005}, the dimensionless flow rate and the maximum velocity respectively scale with $(W/d)^{5/2}$ and $(W/d)^{3/2}$ whatever the angle of the flow. 
Yet, their dependence with $h/W$ confirms the existence of two regimes. At low angles, in regime I, in agreement with the results reported by \cite{Jop_JFM_2005},  $\Qadim$ and $V_{max}/\sqrt{gd}$ scale respectively as $(h/W)^{7/2}$ and $(h/W)^{5/2}$. 
For larger angles, in regime II, $\Qadim$  and $V_{max}$  scale as $(h/W)^2$ and $h/W$, respectively~\citep{Richard2008}. 
We have seen before that $V_{max}/\sqrt{gd}$ scales as $l/W$ in regime I as in regime II, this shows that $h/W$ scales as $(l/W)^{2/5}$ in regime I, 
and thus, $\Qadim$ scales as $(l/W)^{7/5}$.
\revPRb{The difference of scaling for the two regimes can also be confirmed without using $h$ or $l$}, for example, by plotting $V_{max} d / (W\sqrt{gW})$ versus $\Qadim/(W/d)^{5/2}$ (see figure~\ref{fig:tan_Q_Vmax_vs_hsurW}d).
\revPRb{In agreement, with previous results, we indeed recover the existence of different scaling exponents for the two regimes, {i.e.},
$V_{max} d / (W\sqrt{gW})$ scales with $\left[\Qadim/(W/d)^{5/2}\right]^{5/7}$ in regime I and with $\left[\Qadim/(W/d)^{5/2}\right]^{1/2}$ in regime II. The values of the latter exponents are consistent with the scaling laws obtained for $V_{max}$ and $\Qadim$ versus $h/W$.}

\begin{figure}
\begin{center}
\resizebox{0.85\columnwidth}{!}{\includegraphics*{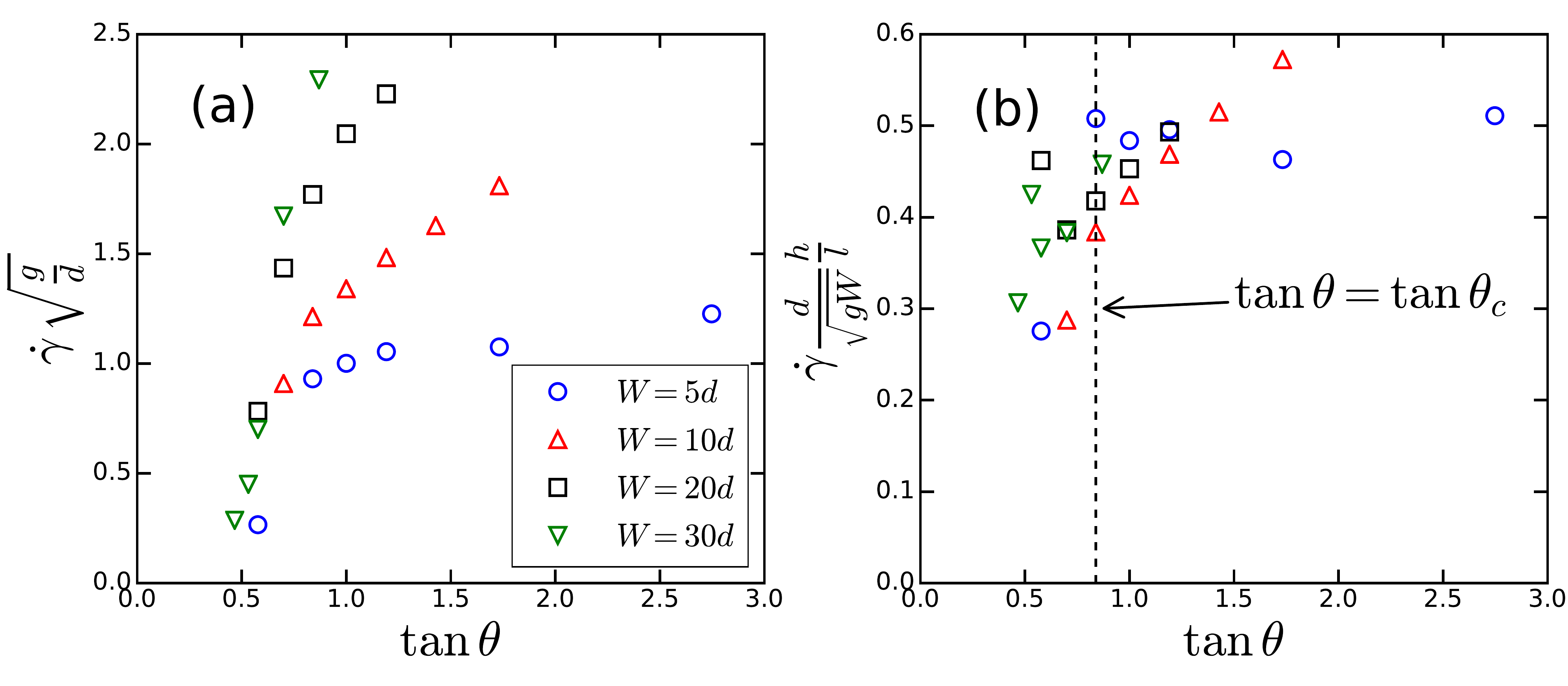}}
\caption{(a) The strain rate increases with the gap between the sidewalls and with the angle of the flow whatever the gap width. Yet, for angles greater that $\theta_c\approx 40^\circ$,  the increase is reduced, confirming the existence of two regimes : $\theta < \theta_c$ and  $\theta >\theta_c$. (b) The data can be
\revPRb{approximately} rescaled on a single master curve by rescaling the velocity by $\sqrt{gW}$. 
}
\label{fig:gammadot}       
\end{center}
\end{figure}

The velocity profiles are not linear both in the creeping and the dilute zones. Consequently the strain rate $\dot\gamma =\partial v_x / \partial z$ depends on $z$. Yet, the velocity profiles have a nearly linear part in the vicinity of $z\approx 0$, region which corresponds to a change of concavity for the shape of the profiles. In the remainder of the paper, we will estimate the strain rate 
at the depth for which the concavity of the vertical velocity profile changes
({i.e.}, at the depth $z$ for which  the strain rate is maximal corresponding to $z/d\approx 0$ for $\theta>\theta_c$).
Figure~\ref{fig:gammadot}a shows the evolution of the strain rate with the angle of the flow and for several cell widths $W/d$. For any value of $W/d$, the strain rate is found to increase with the flow angle. 
For $\theta>\theta_c\approx 40^\circ$ the increase weakens and the slope of $\dot\gamma$ versus $\tan\theta$ decreases with decreasing $W/d$, illustrating once more the existence of the two regimes.
\revPR{When plotting $\dot\gamma d h /  \sqrt{gW} l$ versus $\tan\,\theta$ a reasonable collapse is obtained (figure~\ref{fig:gammadot}b).}
In the creeping zone, there is experimental evidence~\citep{Komatsu_PRL_2001,Crassous_JSTAT_2008,Richard2008,Martinez_PRE_2016} that the mean velocity
of the grains decays exponentially with depth and
the characteristic decay length is approximately equal to
the particle size. The dependence  of this length with the flow rate (and thus the angle of the flow) and the gap width  remains unclear despite some experimental attempts~\citep{Richard2008}.
\begin{figure}
\begin{center}
\includegraphics*[width=0.85\columnwidth]{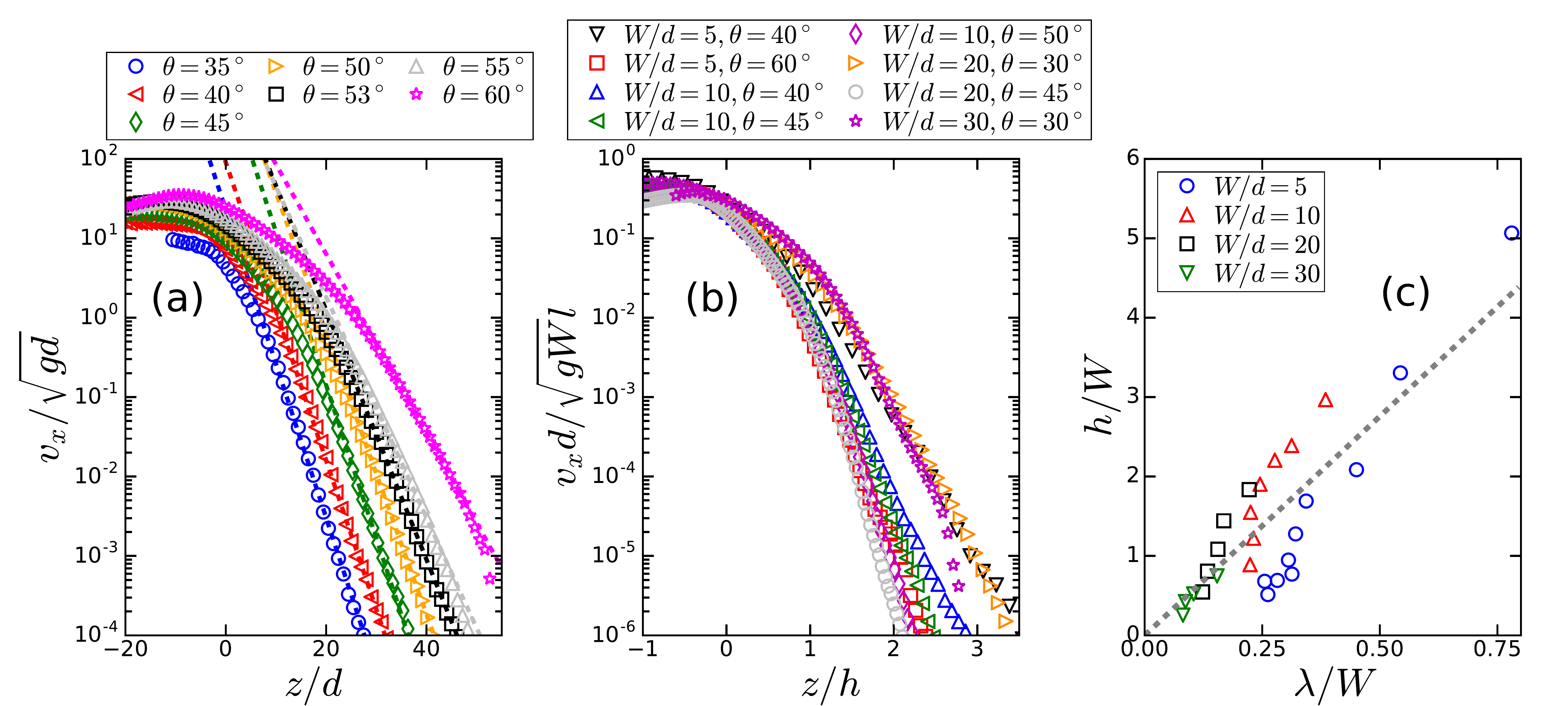}
\caption{In the creeping zone, the velocity decays exponentially with the depth. This is illustrated in  panel (a) for $W/d=10$. 
(b) Rescaling the velocity by $\sqrt{gW}l/d$ and the depth by $h$ leads to a collapse of the velocity profiles in the creeping region 
for $\theta > \theta_c$ (b).
The characteristic length of this decay, $\lambda$, is found to be correlated to the length $h$ (c).
}
\label{fig:creep}       
\end{center}
\end{figure}
Our numerical simulations indeed capture the exponential decay of the grain velocity in the creeping zone (figure~\ref{fig:creep}) over several decades. We found that the characteristic length of the decay, $\lambda$, which is of the order of a grain size, depends on both the gap width and the inclination angle. 
It should be pointed out that our range of flow rates is much more important than those used in the literature~\citep{Komatsu_PRL_2001,Crassous_JSTAT_2008}.
Combining volume fraction measurements by $\gamma$-ray adsorption and imagery, \cite{Richard2008} have shown that both the height of the flowing zone ({i.e.} in our case $h$) and the characteristic length $\lambda$ are linearly correlated.
Our simulations capture this linear relation for both regimes and the slope obtained ($\approx 6$) is close to that reported in~\citep{Richard2008}.
\subsection{Flux densities}
Another quantity of interest is the flux density defined as the product of the volume fraction and the streamwise velocity : 
$\phi(z)=\nu(z)\,v_x(z)$. 
\begin{figure}
\begin{center}
\resizebox{0.85\columnwidth}{!}{\includegraphics*{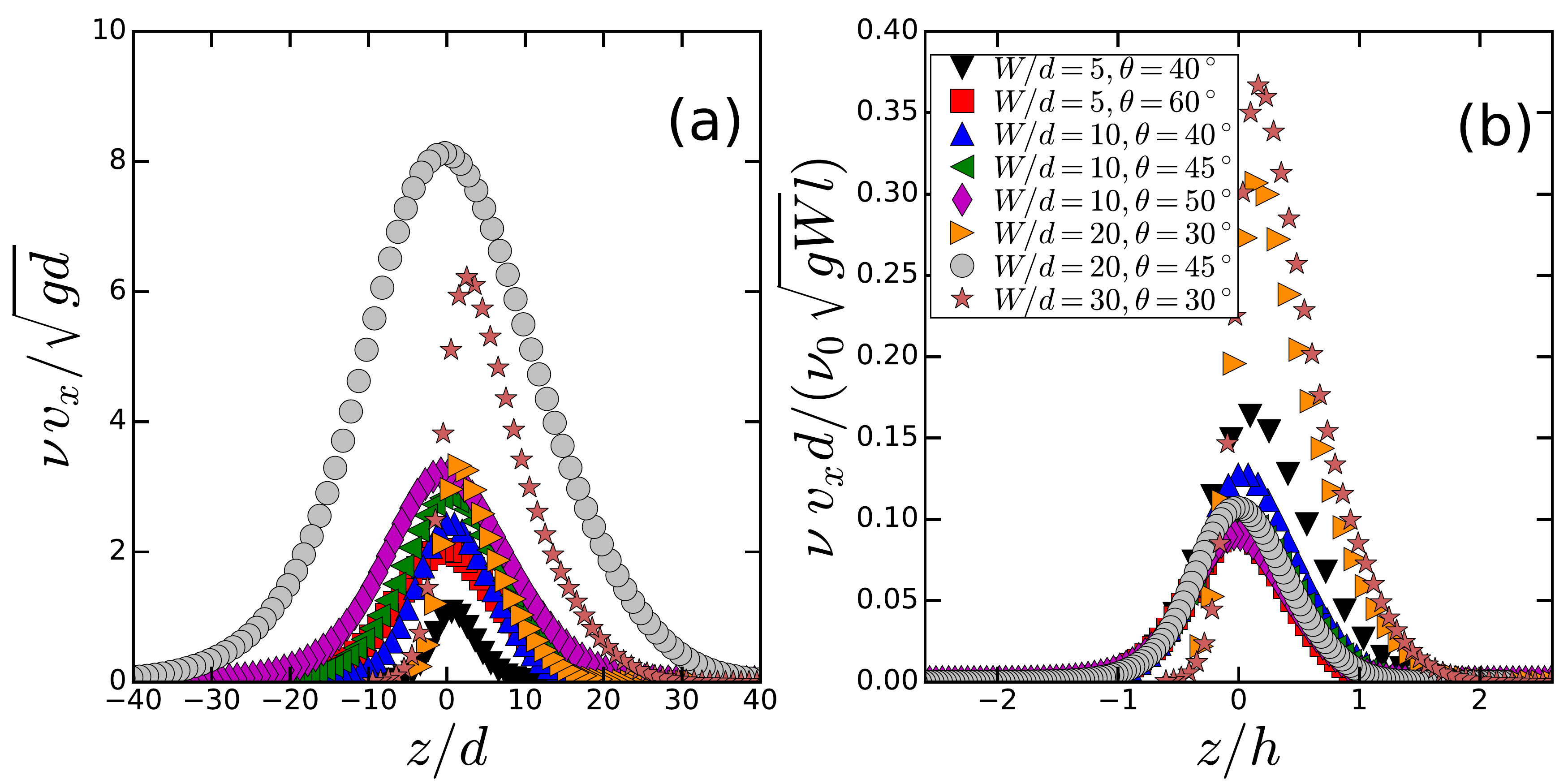}}
\caption{(a) In regime II, the flux densities are maximum at $z\approx 0$.
(b) For the same regime, the vertical profiles of the flux densities collapse on a single curve if the velocity and the depth are respectively scaled by $\sqrt{g W} l/d$ and $h$.
}
\label{fig:flux}       
\end{center}
\end{figure}
It quantifies the number of grains crossing a unit area whose normal is along the streamwise direction.
This quantity is widely used in the context of aquatic and aeolian sediment transport but much less commonly in the physics of granular media.

Figure~\ref{fig:flux}a, which depicts the vertical profiles of this quantity, 
shows that in regime II, the flux densities are almost symmetric with respect to $z/d=0$ {i.e.} $\nu=\nu_0/2$, where they are maximum.
Consequently, contributions to the flow rate of the low density ($\nu<\nu_0/2$) and of the high density region ($\nu>\nu_0/2$) are similar, the former being more rapid than the latter. We have seen that the characteristic lengths for the variation of the volume fraction ($l$) and  the velocity ($h$) are close to each other in regime II. Consequently, the flux density being the product of the volume fraction by the velocity, $h$ (or equivalently $l$) is the relevant length to describe the variation of this quantity in regime II (figure~\ref{fig:flux}b).

We can approximate, in the vicinity of $z=0$, the volume fraction profile by  $\nu(z)\approx \nu_0/2 [1+2z/l]$ and the velocity by $v_x(z)/ \left(\sqrt{gW}l/d\right)\approx \kappa(1-z/h)$ with $\kappa=\dot\gamma d/\sqrt{gW}\approx 0.4$ (where $\gamma$ is calculated at $z=0$). Note that the value of $\kappa$ is consistent with the values of $\dot \gamma d /\sqrt{gW}$ reported in figure~\ref{fig:gammadot}b.
Using the two latter expressions and the approximation $h\approx l$ in regime II allows us to approximate the flux density $\phi(z)/\left(\nu_0 \sqrt{gW}l/d \right)$ by the parabola $\nu_0\kappa\left(1-2z^2/h^2\right)$ in the vicinity of $z=0$. 
This parabola-shaped profile is characteristic of heap flows at large angles. Note that the contribution of the flowing layer ($z\in[-l/2,l/2]$) to the global flux is the most important.  The contribution of the other zones is, however, far from negligible ($\approx 1/3$ of the global flux).

For angles below $\theta_c$, the velocity profile is not symmetric with respect to $z/d=0$, and the length $h$ characterizing the height of the flowing layer is larger than the length $l$ characterizing the variation of the packing fraction with depth. The maximum of the flux density is shifted towards higher volume fractions. 
The volume fraction variation is indeed more localized close to $z/d=0$ and tends toward a step function. We thus recover the behaviour observed 
for SFD dense flows on a bumpy bottom~\citep{Silbert2001,Bi2005,Bi2006} for which the volume fraction is almost constant and, consequently, the flux density is maximum when the velocity is maximum.\\


\subsection{Rotation}
\begin{figure}
\begin{center}
\resizebox{0.75\columnwidth}{!}{\includegraphics*{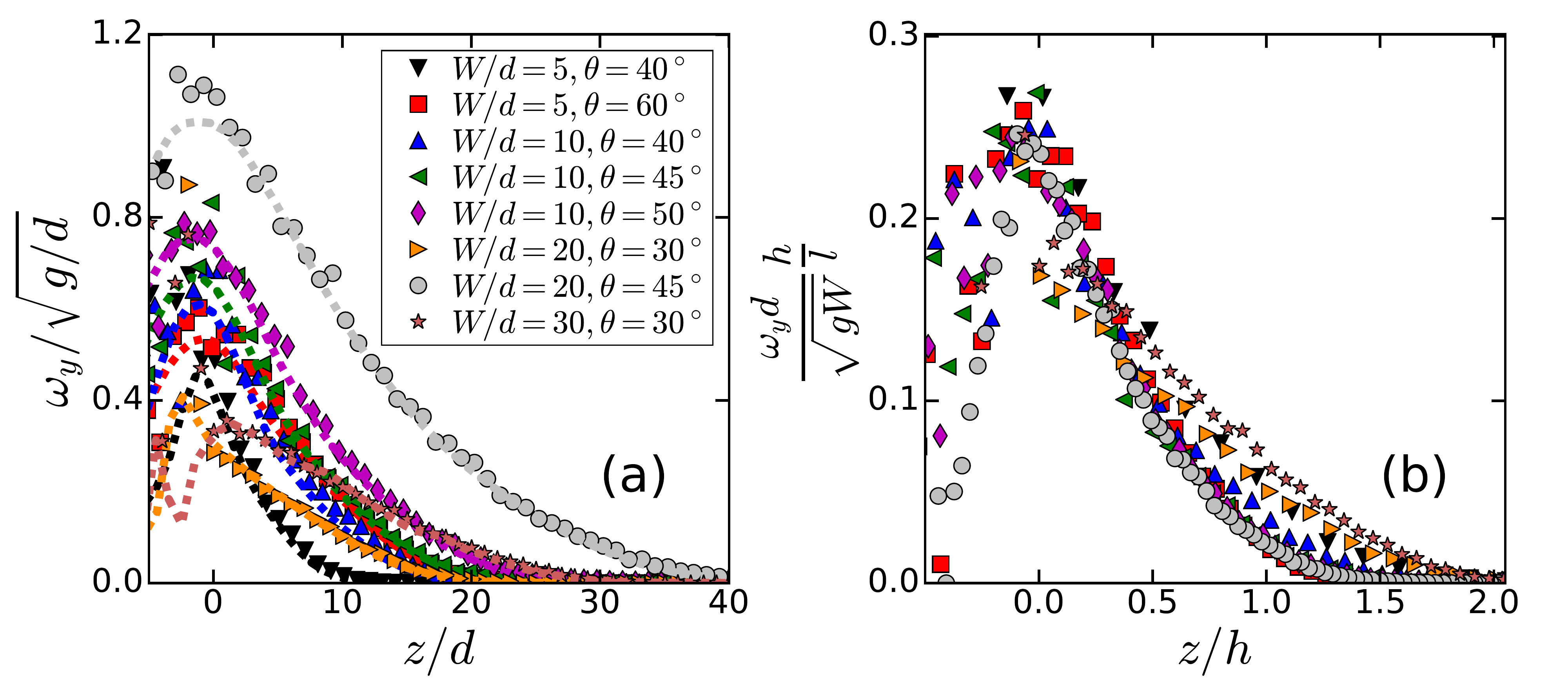}}
\caption{(a) Vertical profiles of rotation velocity in the $y$ direction   for various inclination angles $\theta$ and widths $W$.
The dashed lines correspond to vorticity 
$- 0.5 \partial v_x / \partial z$.
 (b) The scaling with $ d h/ (l \sqrt{gW}$) holds for regime II.
}
\label{fig:profil_omegay}       
\end{center}
\end{figure}

We have also studied the rotation of the grains within the flow which is a quantity of interest since, 
in confined systems, geometrical frustration of rotation consequence of geometrical constraints are frequent. This frustration of
rotation has an important influence on the organization of
the particles’ rotations and thus on the rheology of the system~\citep{Yang_granularmatter_2016}.
The mean angular velocity $\omega_y$ ({i.e.} normal to the sidewalls) profiles  are reported in figure~\ref{fig:profil_omegay}a. 
They are counted positive in the
counter-clockwise direction.
Kinetic theories~\citep{Lun1987,Jenkins_PoF_1985} assume that the mean
angular velocity can be approximated by the rotation rate across 
the flow depth: $\omega_y \approx - 0.5 \partial v_x / \partial z = 0.5 \dot\gamma$. 
In our system, this assumption is 
valid from the creeping zone up to $z/d\approx 0$, {i.e.}, for $z/d>0$ and a volume fraction above $\nu_0/2$.
\revPR{In contrast, the agreement between the two quantities is poorer for $z/d<0$, {i.e.}, for dilute flows.
This discrepancy, which seems to be independent of $\theta$ and $W$, can be explained by the fact that for $z/d>0$ the average number of contacts per grain is almost zero (see section\ref{appA}).  
The probability of collision with sidewalls becomes more significant as $z$ decreases. 
Consequently their velocity distributions are not classical anymore and the relation between $\dot\gamma$ and $\partial v_x / \partial z$ does not hold for these grains.}
Due to the correspondence between $\omega_y$ and $\dot\gamma$, it is natural to represent the vertical profile of $\omega_y$ by scaling the latter quantity by $\sqrt{gW} l / hd$ and $z$ by $h$ (see figure~\ref{fig:profil_omegay}b). As expected, the regimes I and II are recovered due to the link between $\dot\gamma$ and $\omega_y$.\\

\subsection{Temperature profile}
The  velocity fluctuations are characterized by the granular temperature, which is defined by
$T = \left(T_{xx}+T_{yy}+T_{zz}\right)/3$, where $T_{ij}=\langle v_iv_j \rangle - \langle v_i \rangle \langle v_j \rangle$, with $i, j = \{x, y, z\}$.
The granular temperature is a measure
of the agitation of the grains.
One of the main applications of the concept of granular temperature is the derivation of kinetic and hydrodynamic theories for granular systems. It is thus an important quantity that deserves to be studied. 
\begin{figure}
\begin{center}
\resizebox{0.90\columnwidth}{!}{\includegraphics*{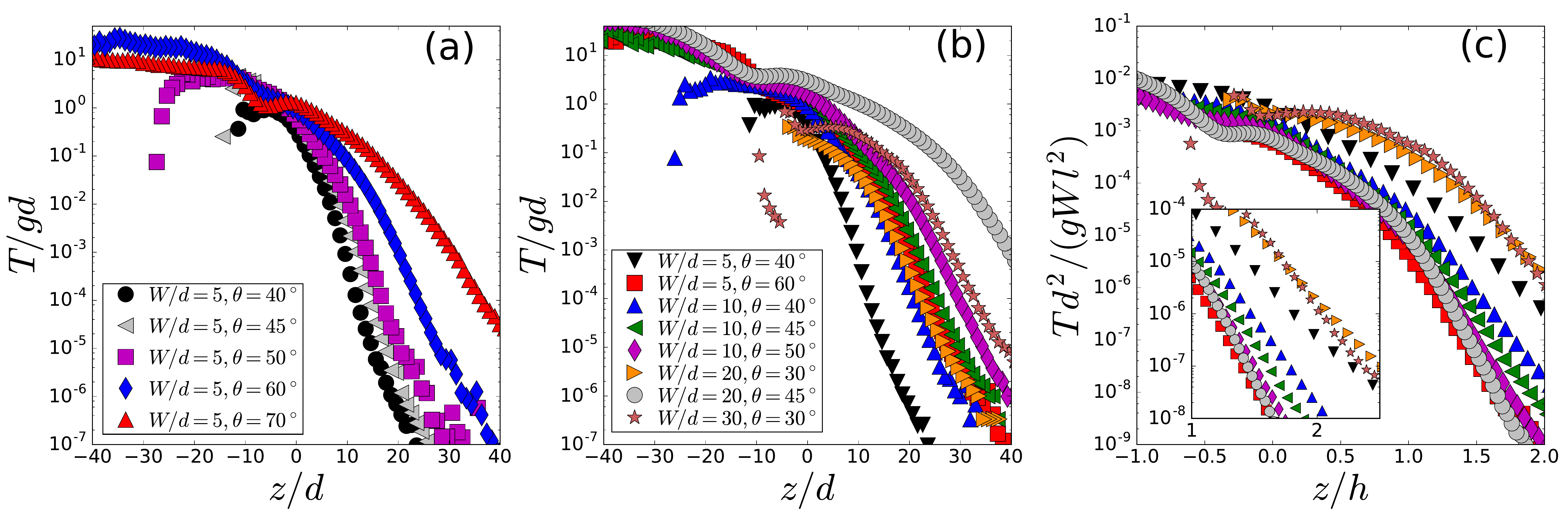}}   
\caption{(a) For a given channel width (here $W=5d$) the vertical profiles of fluctuations of the  particle velocity  increase with $\theta$ the angle of the flow.
The vertical profiles plotted for various widths and angles 
 can be rescaled using $l$ as a length scale and $h$ as a depth scale similarly to what is observed for the velocities (c). A zoom  corresponding to low granular temperatures is reported in the inset of panel (c).
}
\label{fig:profil_T}       
\end{center}
\end{figure}
Figure~\ref{fig:profil_T}a reports the vertical profile of  the granular temperature for several angles of inclination and $W/d=5$. 
The granular temperature is found to increase from the creeping zone (which behaves like a heat sink) to the flowing region. 
More details on the temperature can be found in section~\ref{sec:spanwise_temp} where its properties along the transverse direction are reported. 
Note that the temperature profiles reported here are highly different from those obtained for SFD flows on a bumpy bottom, which behaves like a heat source~\citep{Silbert2001}. Consequently, in the latter case, the granular temperature is maximum in the vicinity of the bumpy bottom~\citep{Silbert2001,Louge_PRE_2003}.
It should be pointed out that the quantities $T_{xx}$, $T_{yy}$ and $T_{zz}$ behave similarly to $T$ ({i.e.} they increase from the creeping region to the free surface). Note that  $T_{yy}\approx T_{zz}$ 
and that $T_{xx}$ is always larger than $T_{zz}$ and $T_{yy}$ ($T_{xx}/T_{yy}\approx 1$ in the creeping region, $\approx 3$ in the flowing zone and up to $10$ close to the free surface).
Similarly to what has been observed in the case of the streamwise velocity profile $v_{x}(z)$, the temperature profiles can be rescaled using $g W l^2/d^2$ as a temperature scale, and $h$ as a depth scale (figure~\ref{fig:profil_T}c). \\


The kinetic theory for fast flows~\cite{Haff_JRheol_1986} assumes that fluctuations
should depend on local relative motion and, consequently,  predicts $T_{xx}\propto \dot\gamma^\chi$ with $\chi=2$. This prediction 
 has been experimentally verified for dilute and rapid chute flows~\citep{Azanza1999}. Yet this scaling is no longer valid in the case of denser and slower flows where an exponent $\chi$ close to $1$ has been reported~\citep{Losert_PRL_2000,Mueth_PRE_2003}, a signature of non-local effects~\citep{Jenkins_granularmatter_2007,Kamrin_SoftMatter_2015}. In a geometry quite similar to ours, \cite{Zhang_EPJE_2019} have recently reported a relation $T\propto\dot\gamma^\chi$ with $\chi\approx1.57$ in the creeping zone. Similarly, \cite{Artoni_PRL_2015} also reported data consistent with $\chi\approx3/2$ for confined shear flows subjected to a constant pressure, which are quite similar to the flows studied in the present work.
In the case of surface flows in a rotating cylinder~\citep{Orpe_JFM_2007}, which are steady but not fully developed, the exponent $\chi$ increases with  the rotational speed of the cylinder  from $\chi=1$ to $\chi=2$.
\begin{figure}
\begin{center}
\resizebox{0.75\columnwidth}{!}{\includegraphics*{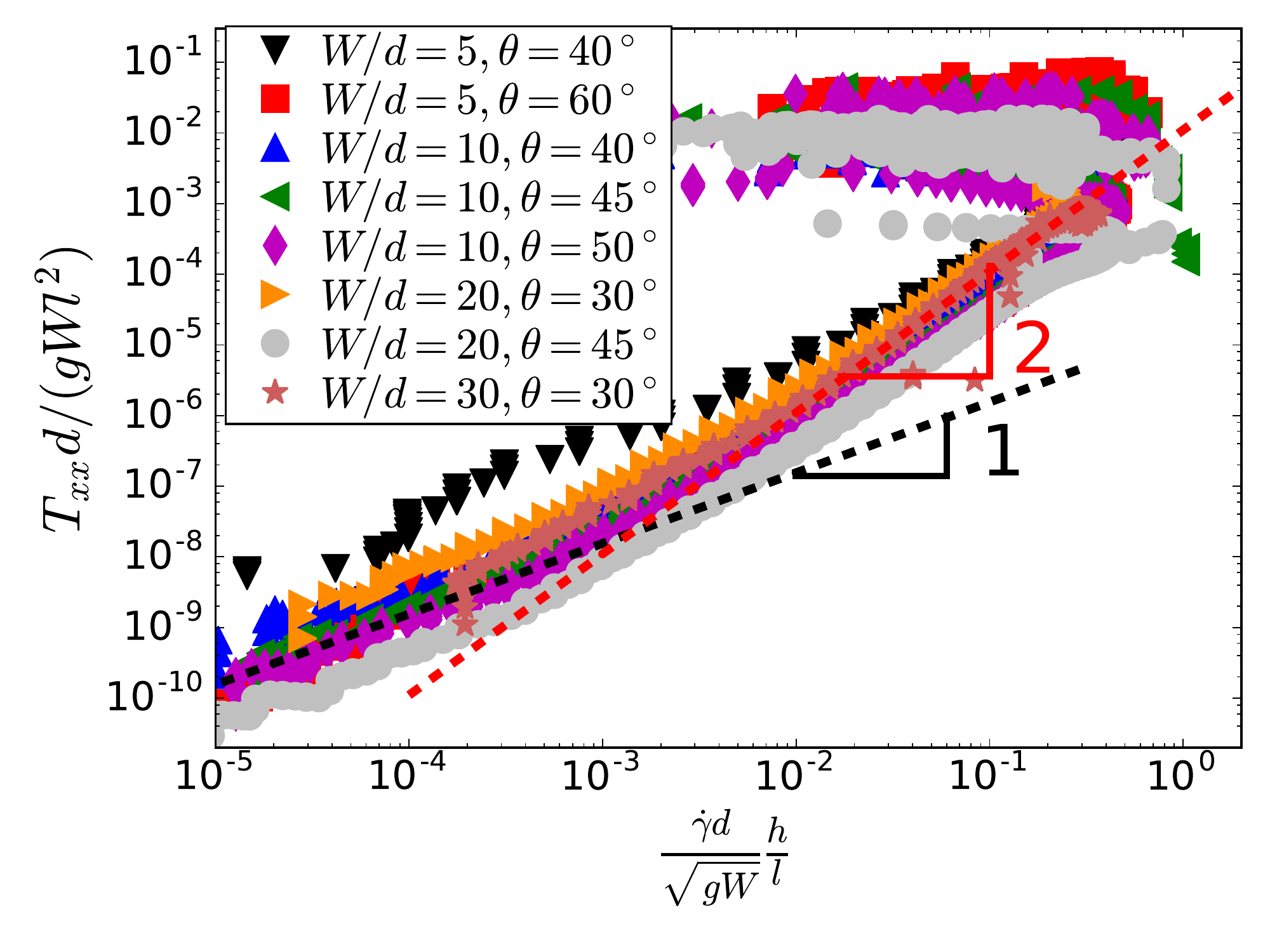}}
\caption{Variation of the dimensionless temperature
$T_{xx} d / g W l^2$ with the strain rate ${\dot\gamma  h d}/{\sqrt{gW}} {l}$ in the flowing layers. The dashed black line shows that the relationship can be approximated by a power law of exponent $\chi = 1$ at the bottom the creeping region and by $\chi=2$ in the buffer and the flowing zones. The whole curve (excepting the part corresponding to the gaseous zone) can be fitted in a satisfactory way by a power law of exponent $\chi=1.5$ (not shown).
}\label{fig:scalingchinois}
\end{center}
\end{figure}
Figure~\ref{fig:scalingchinois} shows the variation of $T_{xx}$ with the strain rate. The reported relation between the latter quantities is consistent with a power law $T_{xx}\propto \dot\gamma^\chi$ but with $\chi$ varying between $\chi \approx 1$ (in the bottom of the creeping zone) and $\chi\approx 2$ (in the buffer and the flowing zones).  In agreement with the literature~\citep{Zhang_EPJE_2019,Artoni_PRL_2015}, if the whole curve (with the exception of the part corresponding to the gaseous zone) is fitted by a power law,  an exponent $\chi$ close to $1.5$ is found.
\revPR{
Remarkably, the variation range of $\chi$ 
is independent of both the
inclination angle $\theta$ and the gap between sidewalls $W$.
This result is in agreement with that obtained in the case of flows in rotating drums~\citep{Orpe_JFM_2007} for which we also have $T_{xx} \propto \dot \gamma^\chi$ with $\chi\in[1,2]$.}
%

\section{Kinematic properties in the transverse direction}\label{sec:transverse}
As mentioned in section~\ref{sec:DEM}, the vertical profiles are averaged over $y$, ({i.e.} along the transverse direction) as classically done in the literature. Yet, due to the presence of sidewalls this averaging may hide the flow behaviour in the vicinity of the sidewalls. They indeed modify the local arrangement of grains near their location and the dissipation in their vicinity is also different from that in the bulk of the flow.   
For this reason, we will report below the transverse evolution of the following quantities: volume fraction, velocity, vorticity and granular temperature. 
\subsection{Volume fraction}\label{sec:compatransverse}
Since the middle of the preceding century~\citep{Verman_Nature_1946,Brown_Nature_1946}, it has been well known that 
the presence of sidewalls influences the volume fraction of static granular materials in their vicinity (see~\citep{Camenen_PRE_2012} and references therein). More precisely, they lead to a reduction in volume fraction due to the wall-induced structure which extends within the packing.
This result can be understood with a simple
description incorporating two regions: (a) an effective boundary layer (which extends from each sidewall to a given length $\Delta_\nu$) and (b) a bulk-like region. 
 The latter region
is assumed to have a volume fraction equal to that of an infinite system whereas the volume fraction of the former region is lower.

\revPR{In our system, a lower volume fraction in the vicinity of the sidewalls is still expected. The creeping zone is not far from being static, the geometrical constraint induced by the latter remains important.
In the flowing zone, as it will be shown in section~\ref{sec:spanwise_temp}, the transverse profile of the temperature  imply a depletion of grains in the vicinity of sidewalls.
We use a description similar to the static modeling with an effective boundary layer of size  ($\Delta_\nu$) depending on the depth within the packing.}

\begin{figure}
\begin{center}
\resizebox{0.75\columnwidth}{!}{\includegraphics*{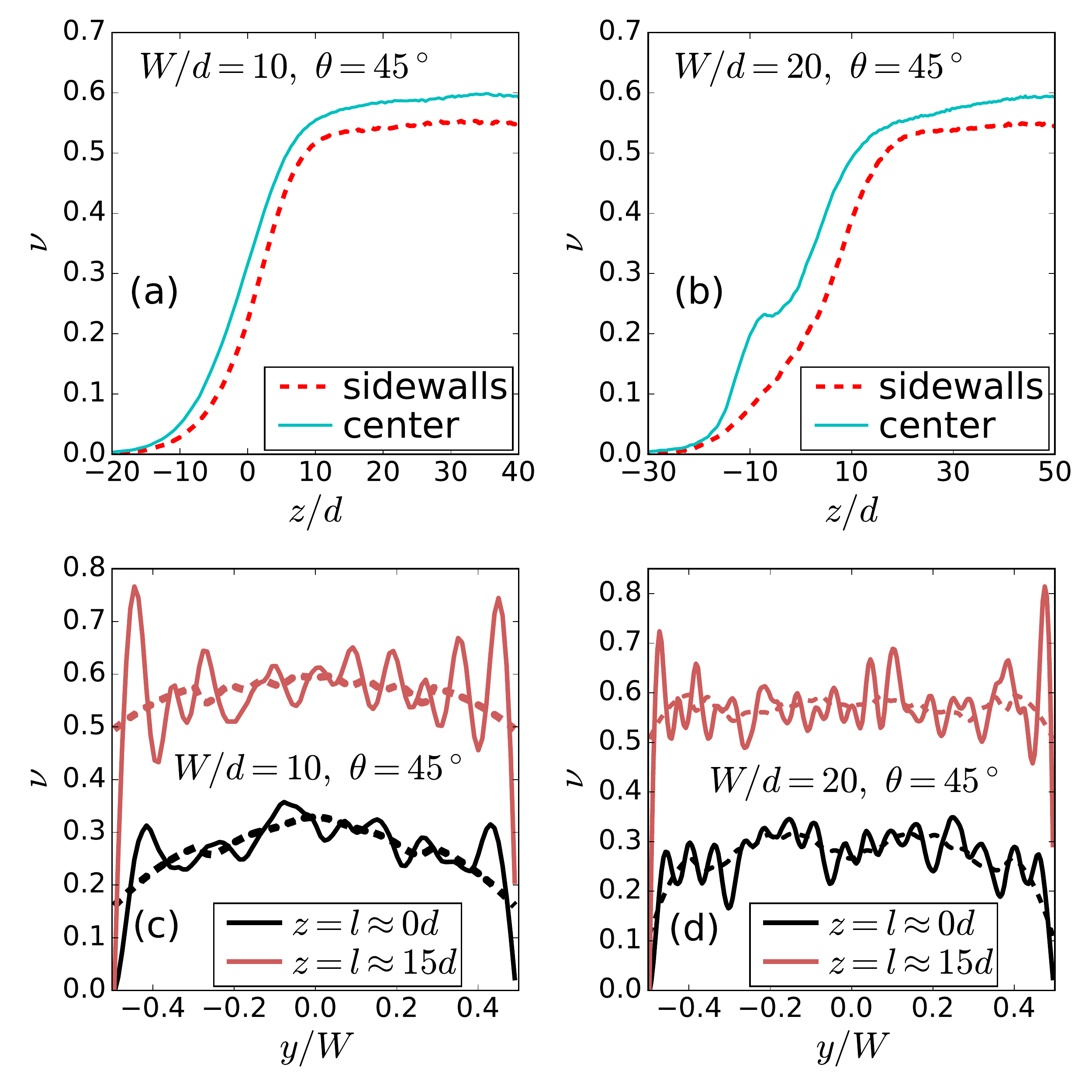}}
\caption{Effect of the sidewalls on the volume fraction profiles for $\theta=45^\circ$.
At any depth within the flow for $\theta=45^\circ$, and for any value of $W$, the volume fraction is smaller in the vicinity of the sidewalls with respect to that measured in the centre of the cell ((a) and  (b) for respectively $W/d=10$ and $W/d=20$).  The effect of the walls is visible on the transverse volume fraction profiles especially in the flow region ((c) and (d) for respectively $W/d=10$ and $W/d=20$). 
The solid lines correspond to data averaged over a length of $0.1d$ and the dashed lines correspond to the same data filtered by a Savitzky-Golay filter (window length $4d$ and polynomial degree $3$).
The volume fraction can be considered as uniform only for large $W/d$ and far from the sidewalls.
}
\label{fig:compa3D}       
\end{center}
\end{figure}

We have  reported in figure~\ref{fig:compa3D}
the transverse volume fraction profiles for two flow depths, one in the flowing region ($z/d=0$) and the other one in the creeping zone, for a flow angle equal to $45^\circ$ with $W=10d$ (figure~\ref{fig:compa3D}(c)) and $W=20d$ (figure~\ref{fig:compa3D}(d)). The solid lines correspond to data averaged over a length of $0.1d$. They clearly reveal the layering induced by the flat walls. It is especially important in the creeping zone, but does not disappear completely in the flowing zone.To get free from the fluctuations induced by this layering, we filtered the data to produce smoothed volume fraction profiles corresponding to the dashed lines of the figure. 
The presence of sidewalls clearly leads to a decrease of the values of the smoothed volume fraction in their vicinity. This effect is stronger in the flowing zone that in the creeping zone. It should be pointed out  
that even a small modification of the volume fraction has important consequences for the global macroscopic behaviour of a granular system.
In the creeping zone, the value of $\Delta_\nu$ seems independent of $W$ and is of the order of 3 or $4d$. For $W=10d$ the volume fraction in the flowing region ($z/d=0$) is strongly influenced by the sidewalls:  it increases continuously from the sidewalls to the centre of the cell. In contrast, for $W=20d$, also for $z/d=0$, a plateau of volume fraction seems to be reached between $y=-W/4$ and $y=W/4$.

We have  reported in figure~\ref{fig:compa3D}(a) and (b) the vertical profiles of the smoothed volume fraction in the middle of the cell and at the sidewalls for the same configurations. As expected, the volume fraction is always smaller in the vicinity of the sidewalls. This also confirms that this effect is stronger in the flowing zone.

It should be pointed out that, for $W=20d$ and $\theta=45^\circ$, \revPR{around $z/d=-10$}, an unexpected increase of the volume fraction (a bump) is observed in the centre of the cell in the vicinity of the free surface. This is due to the formation of dense clusters of a few grains. The exact characterization of these clusters is beyond the scope of the present paper.\\


\subsection{Velocity profiles}
\begin{figure}
\begin{center}
\resizebox{0.75\columnwidth}{!}{\includegraphics*{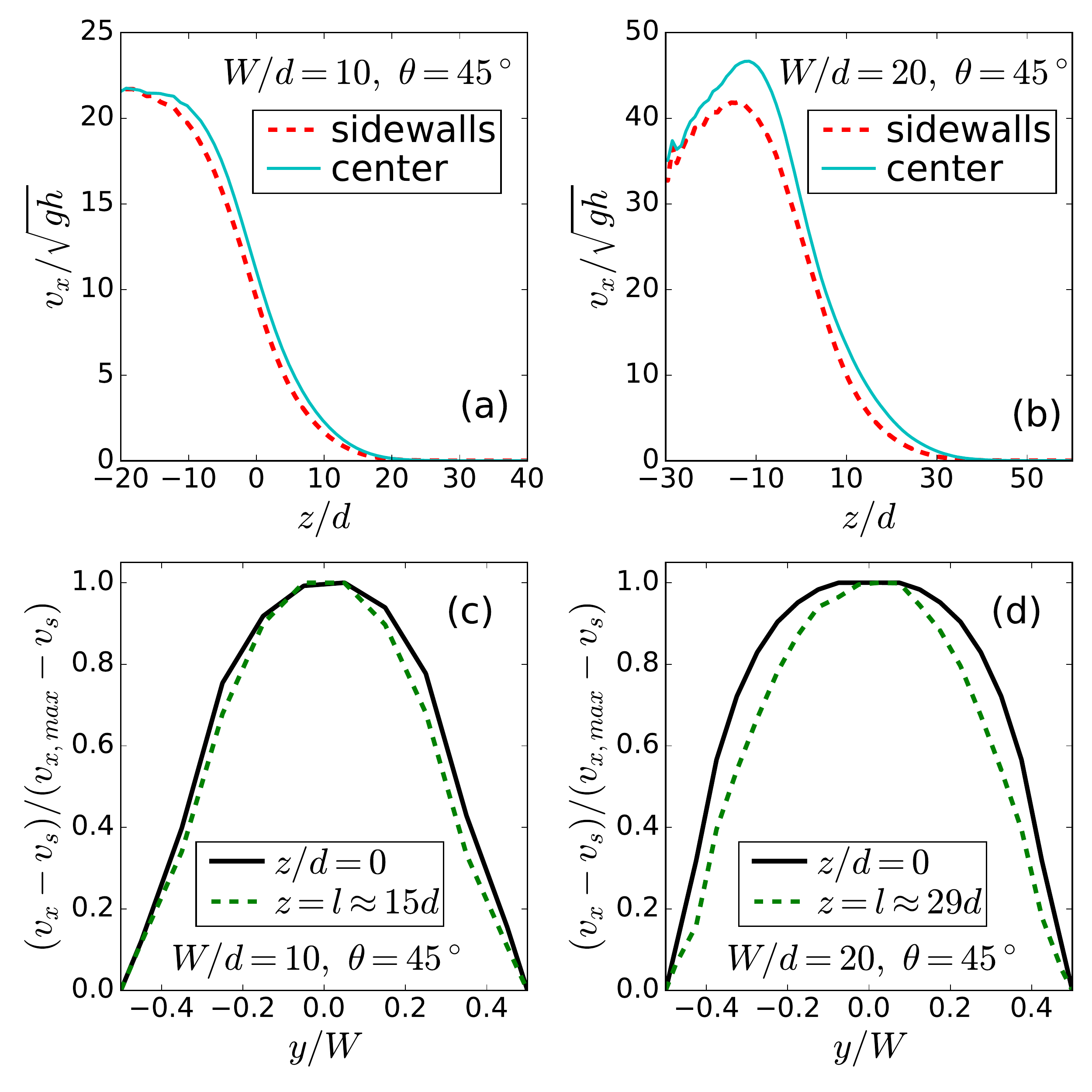}}
\caption{Effect of sidewalls on the velocity profiles for a flow angle equal to $45^\circ$ {i.e.} greater that $\theta_c$, the angle corresponding to the transition between regime I and regime II. 
For small gap width ((a) $W/d=10$) the difference between the vertical velocity profiles at sidewalls and  in the centre of the simulation cell is small and increases with increasing $W$ ((b) $W/d=20$). 
The influence of the sidewalls on the 
transverse profile of the reduced velocity at a given depth $z$ (i.e. $(v_x-v_s)/(v_{x,max}-v_s)$ where $v_{x,max}$ and $v_s$ are respectively the maximum velocity and the sliding velocity at the aforementioned depth) is weak but increases from the flow region to the creeping one. 
This variation with depth also increases with increasing $W$ (c) and (d).}
\label{fig:vit3D_sup}       
\end{center}
\end{figure}

Bumpy sidewalls are generally considered as non-sliding boundaries~\citep{Jop_Nature_2006} with a vanishing velocity at the sidewalls. Here, the sidewalls are flat, the granular material is expected to have an average velocity that is not negligible at their contact. We have reported in figure~\ref{fig:vit3D_sup}(a) and (b) the vertical profiles 
at sidewalls and in the centre of the simulation cell for $\theta=45^\circ>\theta_c$, with $W=10d$ and $W=20d$, respectively. As expected, an important longitudinal velocity at the sidewalls is observed. We call it the sliding velocity and we denote it as $v_s(z)$  at depth $z$. 
Our simulations show that the transverse profiles of velocity are similar in regimes I and II. The nature of the regime has an effect on the evolution of the sliding velocity with the angle, but the behaviour of the reduced velocity at a given depth $z$, defined as  $\left(v_x(y,z)-v_s(z)\right)/\left(v_{x,max}(z)-v_s(z)\right)$, where  $v_{max}({z})$ is the maximum value of $v_x(y,z)$ at depth $z$, independent of the regime.

Transverse profiles of reduced velocity are reported in figures~\ref{fig:vit3D_sup}(c) and (d). They are found to broaden (i) from the creeping region to the flow region and (ii) with $W/d$
(see Figs.\ref{fig:vit3D_sup}(c) and (d)). These results highlight the three-dimensional structure of the flows observed in our system.

The transverse velocity profiles can be fitted in a convenient way by 
$v_x(y,z)=v_{max}(z) - B\left[1 - \cosh(y/\Delta_{v_x}) \right]$ with $B=\left[v_{max}(z) - v_s(z)\right]/\left[1 - \cosh\left(W/2\Delta_{v_x}\right)\right]$.
In the latter expression,  
$\Delta_{v_x}$ is the length characterizing the influence of the sidewalls~\citep{Courrech2003,Zhang_EPJE_2019}. We have determined $v_s$ and $\Delta_{v_x}$ by fitting the transverse profiles of the velocities averaged between the depths $-l/2$ and $l/2$. 
\begin{figure}
\begin{center}
\resizebox{0.75\columnwidth}{!}{\includegraphics*{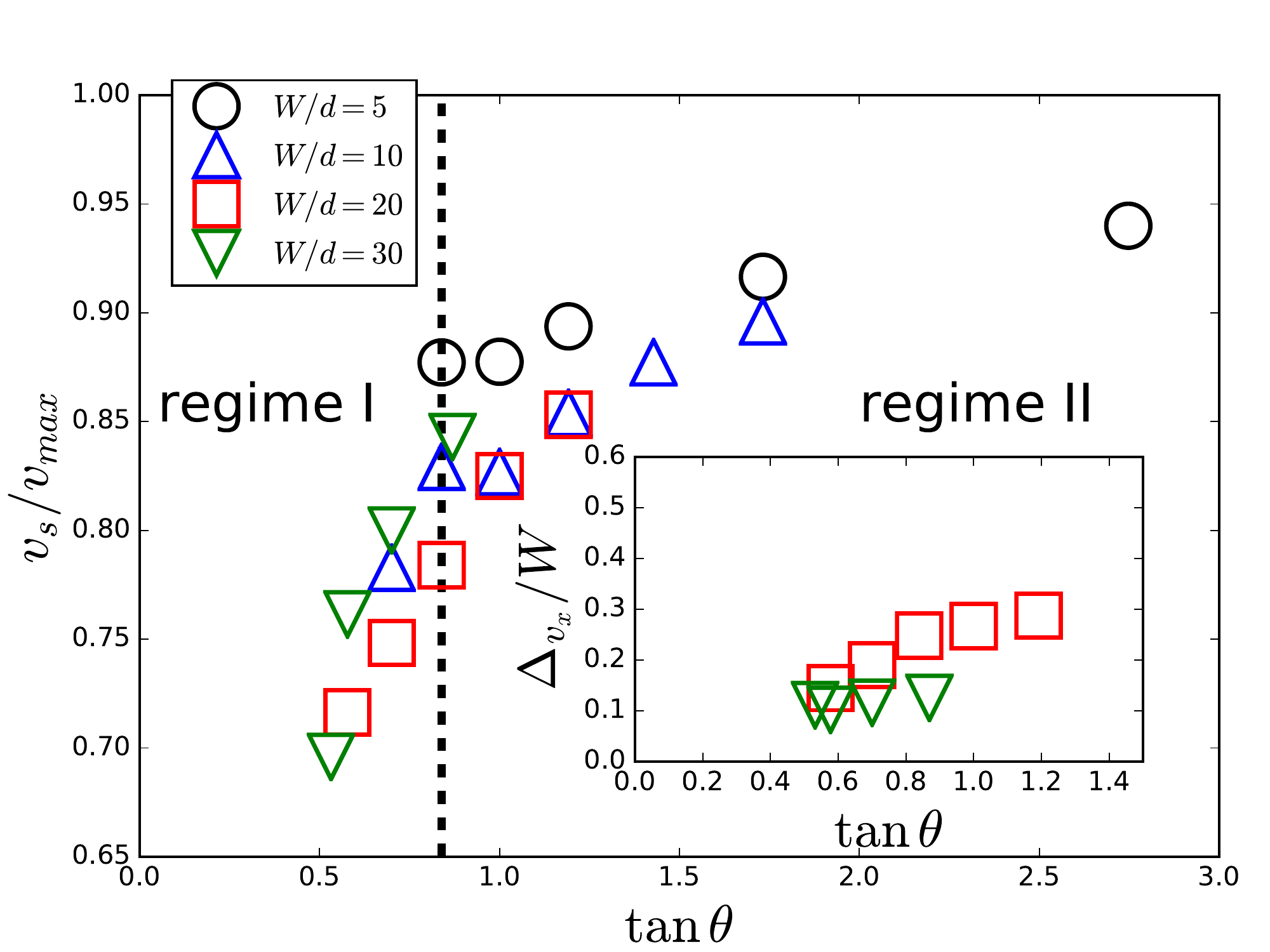}}
\caption{The longitudinal sliding velocity at the sidewalls, $v_s$, increases with the flow angle $\theta$. Interestingly, when rescaled by the maximum velocity, the latter increase is weakly dependent on the gap between the sidewalls $W/d$. As shown on the inset, the length characterizing the influence of the sidewalls (see text) increases  
with $\tan\theta$. 
}\label{fig:vslide}
\end{center}
\end{figure}
The ratio $v_s/v_{max}$ is found to always increase with the flow angle (figure~\ref{fig:vslide}), but it increases much more moderately in regime II than in regime I.
Interestingly, the effect of $W$ on this ratio is weak. 
This seems to indicate that a boundary condition on the rescaled sliding velocity, $v_s/v_{max}$, can be expressed, in a first approximation, as a function of  $\theta$. 

Interestingly, the characteristic length $\Delta_{v_x}$ is much greater than $W/2$ for $W=5d$, indicating that the sidewalls influence the flow across the total width of the channel \revPR{and, consequently, its determination is imprecise.} 
For $W=20d$ and $W=30d$, the characteristic length is smaller than $W/2$ 
(inset of figure~\ref{fig:vslide}) indicating that the effect of the sidewalls is limited to a portion of the channel.
The inset also shows that $\Delta_{v_x}/W$ increases with $\theta$ and decreases with $W$. However, at a given $\theta$, $\Delta_{v_x}$ increases with $W$ (not shown) confirming that, in agreement with~\cite{Jop_JFM_2005}, the effect of the sidewalls is significant even at large gap width. 
The case $W=10d$ is intermediate between the two latter configurations. 
In any cases, the effect of the sidewalls cannot be neglected.
The value of $\Delta_{v_x}$ seems to be of the same order as $\Delta_\nu$, the size of the boundary layer determined from the volume fraction profiles, in the flowing zone.

\subsection{Rotation}
\begin{figure}
\begin{center}
\resizebox{0.75\columnwidth}{!}{\includegraphics*{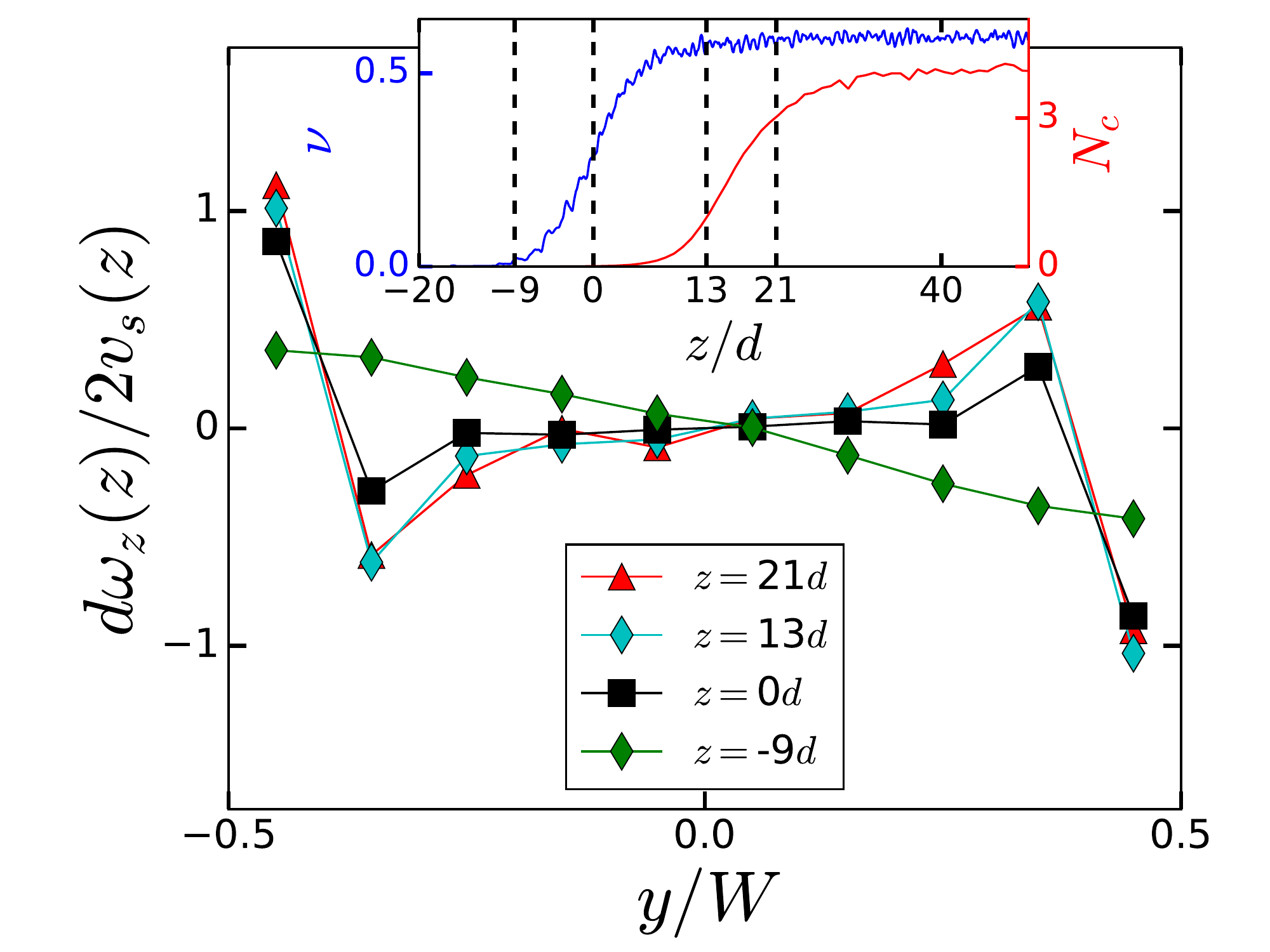}}
\caption{Profiles of vertical rotation velocity rescaled by $2 v_s/d$ versus $y$ for an angle $\theta=40^\circ$ and for several vertical positions. The gap between the sidewalls is $W/d=10$. For the sake of clarity, the inset reports the aforementioned vertical positions (dashed lines) on the vertical profiles of the number of contacts, $N_c$ and of the volume fraction, $\nu$.
}
\label{fig:profil_omegaz}       
\end{center}
\end{figure}

The component of the rotation velocity normal to the bottom ({i.e.} $\omega_z$) is also of interest since it is potentially strongly influenced by the presence of the flat but frictional sidewalls~\citep{Yang_granularmatter_2016}. 
In figure~\ref{fig:profil_omegaz}, we have reported the transverse variations of $\omega_z$ at different flow depths. For the sake of clarity, we have reported in the inset of this figure  the volume fraction and the number of contact profiles and indicate on them the latter depths. Like all the other quantities reported in this paper, this quantity has been 
space and time averaged (see the end of Section~\ref{sec:DEM}). 
In the vicinity of the sidewalls and in the creeping zone, the rotation velocity $\omega_z$ is close to $2 v_s /d$. This indicates that, at these depths, the grains located in the vicinity of the sidewalls mainly roll without sliding at the contact with the sidewall. 
The influence of the sidewalls decreases in the flow zone, indicating that the propensity of the grains for sliding at contact increases. 
The correlations of the rotations along the transverse direction are complex. 
Counter-rotations 
(anti-correlated rotations)  are observed close to the sidewalls (see figure~\ref{fig:profil_omegaz}). The number of involved layers increases with the depth within the flow. It should be pointed out that the observed behaviour is somewhat similar to that of a  system \revPR{with frustrated rotations}~\citep{Radjai_PRE_1996,Khidas_EPL_2000}. 

\subsection{Granular temperature}\label{sec:spanwise_temp} 

\begin{figure}
\begin{center}
\resizebox{0.75\columnwidth}{!}{\includegraphics*{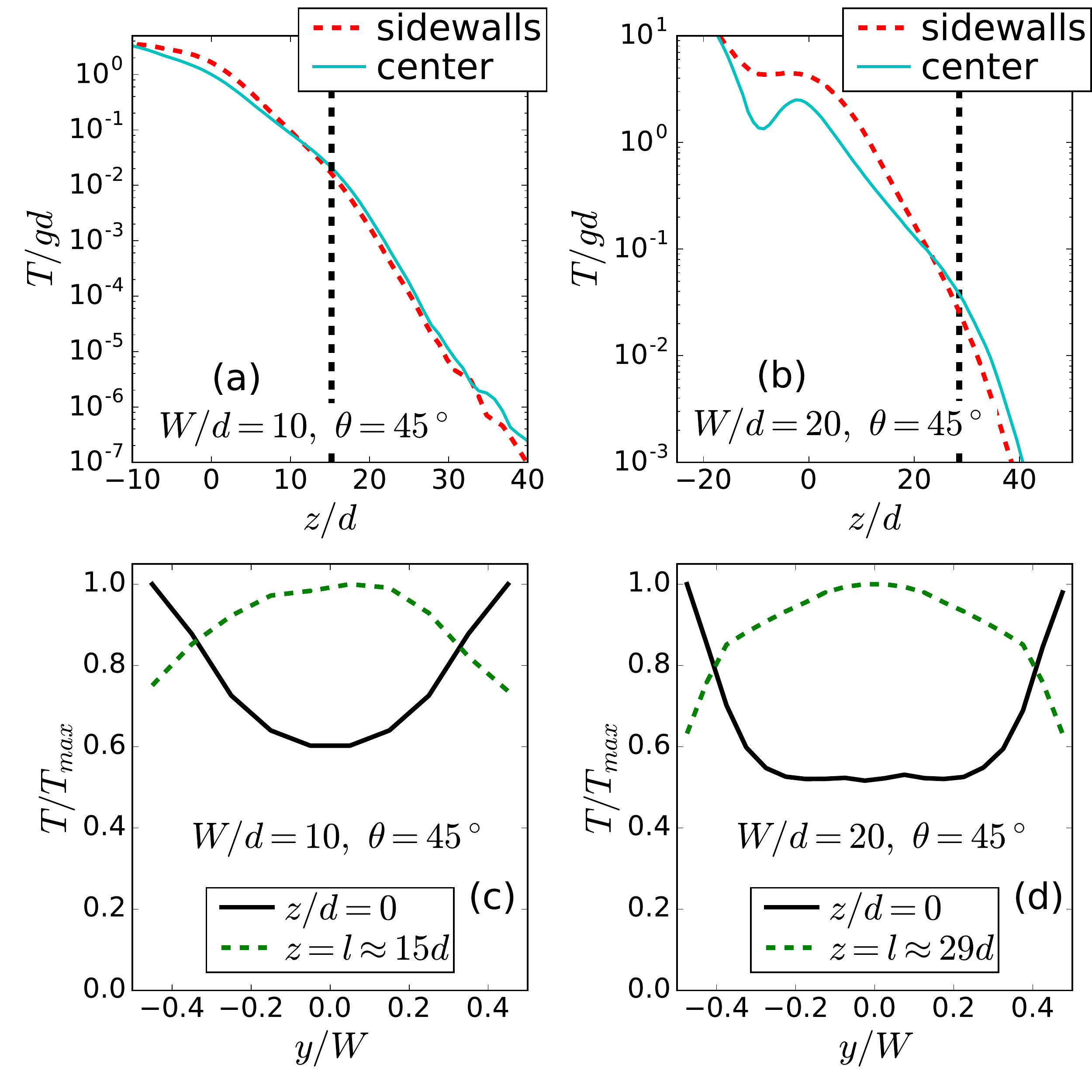}}
\caption{Transverse temperature profiles for $\theta=45^\circ$, $W/d=10$ (a) and   $\theta=45^\circ$, $W/d=20$ (b) in the the centre of the simulation cell and at the sidewalls. The corresponding transverse temperature profiles are respectively reported in (c) and (d). 
The absolute difference between the vertical profiles of temperature at the centre and at the sidewalls is weak in the case of small gap width (a), yet the transverse relative variations are not negligible (c). 
The vertical dashed lines reported in (a) and (b) correspond to $z/d=l$.
In the case of large gap width, the transverse variations of the granular temperature are important, as shown by the vertical profiles at the sidewalls and at the centre (b) as well as on the transverse profiles (d). Interestingly, 
depending on the flow zone (flowing or creeping) the sidewalls behave like a
source or a sink of granular temperature (c) and (d).
}
\label{fig:profil_Txx}       
\end{center}
\end{figure}
At the centre and at the sidewalls the temperature is found to increase from the creeping zone to the free surface, the differences between the two profiles being more important for larger gaps $W$.
\revPR
{Note that for $W/d=20$ a decrease of granular temperature  is observed at $z/d\approx -10$. This location corresponds to the kink observed on the volume fraction profile (see figure~\ref{fig:compa3D}).
As mentioned in section~\ref{sec:compatransverse} it corresponds to the formation of clusters of grains.}
The transverse profile of the  granular temperature is very informative. Outside the creeping zone, this quantity is greatest at the sidewalls and lowest in the centre. In contrast,
in the creeping zone, the granular temperature gradually rises from its minimal value
at the sidewalls to a maximum value at the centre of the cell. 
Note that similar results have been observed in a confined shear cell~\citep{Artoni_JFM_2018,Richard_GM_2020}.
These results have important consequences.
Depending on the vertical position,  sidewalls
could be either a granular heat source (in the flowing zone) or a sink (in the creeping
zone). This points out the difficulty in stipulating a
sidewall boundary condition on the granular temperature for theories aiming to capture
the properties of granular flows involving both creeping and flowing zones.
Note that, in the flowing  zone, the observed high temperature induces an important pressure and a low volume fraction (see section~\ref{sec:compatransverse}). 
\\
A comparison between figure~\ref{fig:profil_Txx} and figure~\ref{fig:vit3D_sup} suggests that the length of influence of the sidewalls on the granular temperature is weaker than that on the velocity. To illustrate this, we can compare the transverse
profile of the temperature and the velocity for $W/d=20$ and $\theta=45^\circ$ ({i.e.}, 
figure~\ref{fig:profil_Txx}d and figure~\ref{fig:vit3D_sup}d,  respectively). For both quantities, a plateau is observed at the centre of the cell, yet that observed for the temperature is significantly larger than that of the velocity. 
In this case, the length characterizing the  effect of the sidewalls on the granular temperature, estimated by the distance between a sidewall and the beginning of the plateau,  is approximately equal to the half of that measured on the streamwise velocity profile.

\section{Conclusion}\label{sec:conclu}
We have investigated confined granular flows over an erodible bed 
by means of DEM. We have studied the kinematic properties of the steady and fully developed regime.  
In contrast to the flow regimes explored experimentally by \citet{Jop_JFM_2005}, who actually limit their investigation to shallow inclination angles (typically between $20$ and $30^\circ$), 
we explored flow regimes up to much larger inclination angles, between approximately $25$ and $70^\circ$. To obtain these steep inclinations, we investigated flow configurations with a small gap $W$ between the sidewalls
($W/d=5,\ 10,\ 20$ and $30$). 

We have characterized in detail the vertical and transverse profiles of the particle volume fraction, particle velocity, particle flux density, particle
rotation speed and granular temperature as functions of the inclination $\theta$ and the gap width $W$. 
This analysis reveals the existence of two distinct regimes at low and large angles of inclination, regimes I
and II, respectively. The transition has been found to occur at a critical inclination angle $\theta_c\approx 40^\circ$. Importantly, two 
different length scales have been identified: the flow height $h$ characterizing the decay of the streamwise velocity profile and the characteristic length $l$ characterizing the decay of the particle volume fraction profile. These two lengths coincide in regime II but differ in regime I, where $(h/W) \propto (l/W)^{2/5}$. 
Taking into account the fact that the experimental results are relative to the length scale $h$,
 we choose to use $h$ to express the scaling with depth. 

In addition to these two length scales,
a unique characteristic velocity scale, $V_c=(l/W)\sqrt{gd}$, was uncovered. The maximum velocity $V_{max}$ scales with this characteristic velocity
$V_c$ both in regimes I and II. Interestingly, the vertical profiles of the velocity and temperature,  when rescaled by $V_c$ and $V_c^2$ respectively, fall
onto a master curve when plotted as a function of the rescaled depth $(z/h)$. We obtain two different distinct master curves according to the flow regime.
Transverse profiles also reveal that the effect of the sidewall extend over a significant part of the width of the chute flow in the range of gap width 
investigated so far ($5d<W<30d$). For $W=5d$, the influence of the
sidewall affects the whole width.


\revPR{We confirm  the robustness of the
relationship between the rescaled flow height $h/W$ and the inclination angle, previously established in \citep{Taberlet2003,Richard2008}:
\begin{equation}
\frac{h}{W} = \frac{\tan \theta- \mu_{b,h}}{\mu_{w,h}} \>,
\end{equation}
where $\mu_{b,h}$ and $\mu_{w,h}$ were interpreted as effective friction coefficients. This relationship 
holds whatever the flow angle and the gap width. A detailed discussion of the establishment and interpretation of this relationship was out of the scope of the present article. This will be one of the goals of the second article of the series.}

Another salient feature is revealed by the simulations. We were able to extract scaling laws for $\dot\gamma$,
$V_{max}$ and $Q$ that are found to depend on the dimensionless length scales $h/W$ and $W/d$.
These scalings support the existence of the two distinct regimes at flow angles smaller and greater than $\theta_c\approx 40^\circ$
(see Table~\ref{scaling_summary}). 
More precisely, in regime II, for a given rescaled flow height $h/W$ (that is for a given inclination), we find the same scalings with respect to $W$
as those derived by \citet{Jop_JFM_2005} : $\dot\gamma \propto W^{1/2}$, $V_{max} \propto W^{3/2}$ and $Q\propto W^{5/2}$.
In contrast, for a given gap width  $W$, the scalings with respect to $h$ obtained in regime II differ from those from \citet{Jop_JFM_2005}. We obtain
$\dot\gamma \propto cst$, $V_{max} \propto h$ and $Q\propto h^2$ whereas \citet{Jop_JFM_2005}
get $\dot\gamma \propto h^{3/2}$, $V_{max} \propto h^{5/2}$ and $Q\propto h^{7/2}$. These contrasting behaviours are not surprising
since flows at small and large inclination angles differ in nature. In the small angle regime investigated by \citet{Jop_JFM_2005}, that is regime I,
the flowing layer remains dense and can be well described by the $\mu(I)$ rheology while, in the large angle regime, 
we observe a strong variation of the volume fraction from the bottom to the top of the flowing layer. This difference is confirmed by the fact
that the relationship between $h$ and $l$ differs in regimes I and II, as emphasized previously. 

\revPR{An important future task  would be to complement the numerical investigations with experiments to confirm the scaling laws in the regime II. We are
currently conducting experiments using a chute flow which was designed to allow for large flow rates. We hope to report soon about these experimental results.}
\begin{table}
\begin{tabular}{c|c||c|c|c|c|c|}
\hline \multicolumn{2}{c||}{Dimensionless quantities} 
& $\frac{\dot\gamma}{\sqrt{g/d}}$ & $\frac{V_{max}}{\sqrt{gd}}$ &  $\frac{\Qadim}{d\sqrt{gd}}$ 
& $\frac{\Qadim/d\sqrt{gd}}{(W/d)^5}$ 
& $\frac{l}{W}$ 
\\
\hline  
Regime I: $\theta < \theta_c$  & Experiments,& & & & &\\
 \ & Simulation 
& $\left( \frac{W}{d} \right)^{1/2} \left( \frac{h}{W}\right)^{3/2}$ & $\left( \frac{W}{d} \right)^{3/2} \left(\frac{h}{W}\right)^{5/2}$ & 
$\left( \frac{W}{d}\right)^{5/2} \left( \frac{h}{W}\right)^{7/2}$ & 
$\left( \frac{V_{max}/\sqrt{gd}}{(W/d)^{3/2}} \right)^{7/5}$
& $\left( \frac{h}{W} \right)^{5/2}$  
\\
& $\&$ Model & &  & & &\\
\hline  \multirow{2}*{Regime II: $\theta > \theta_c$} & Simulations 
&$\left( \frac{W}{d} \right)^{1/2}$ & $\left( \frac{W}{d} \right)^{3/2} \left(\frac{h}{W}\right)$ & 
$\left( \frac{W}{d}\right)^{5/2} \left( \frac{h}{W}\right)^{2}$ & 
$\left( \frac{V_{max}/\sqrt{gd}}{(W/d)^{3/2}} \right)^{2}$
& $\left( \frac{h}{W} \right)$ 
\\ 
\hline 
\end{tabular}
\caption{ Summary of the scaling obtained numerically for $\dot\gamma$, $V_{max}$ and $Q$ 
respectively. We also recall the results obtained by \citet{Jop_JFM_2005} at small angles
derived from the $\mu(I)$ rheology and checked experimentally. Note that the second to last column of the table indicates the scaling
between $\Qadim$ and $V_{max}$ and the last one the scaling of $l/W$ with respect to $h/W$ according to the flow regimes.}
\label{scaling_summary}
\end{table}

\vspace*{1.2cm}\par
\textbf{Acknowledgments.}
Part of the numerical simulations were carried out at the CCIPL
(Centre de Calcul Intensif des Pays de la Loire).  
We also thank Riccardo Artoni for stimulating discussions on confined flows.\\

\textbf{Funding.} P.R. acknowledges the support of the French Research National Agency through the project ANR-20-CE08-0028. R.D, A.V. and P. B. acknowledge the support of the French
Research National Agency through the project ANR-16-CE01-0005.\\

\textbf{Declaration of interests.} The authors report no conflict of interest.\\


\bibliographystyle{jfm}
\bibliography{flowbib}

\end{document}